\documentclass[aps,prb,preprint,onecolumn,citeautoscript]{revtex4-1}  
\usepackage{amsmath,amssymb,mathrsfs,bm,feynmf,setspace}
\usepackage{graphicx} 
\usepackage[tight]{subfigure}   
\usepackage{color} 
\usepackage[papersize={8.5in,11in}]{geometry}
\usepackage[colorlinks=true]{hyperref}   
\hypersetup{
    bookmarks=true,         
    unicode=false,          
    pdftoolbar=true,        
    pdfmenubar=true,        
    pdffitwindow=false,     
    pdfstartview={FitH},    
    pdftitle={My title},    
    pdfauthor={Author},     
    pdfsubject={Subject},   
    pdfcreator={Creator},   
    pdfproducer={Producer}, 
    pdfkeywords={keyword1} {key2} {key3}, 
    pdfnewwindow=true,      
    colorlinks=true,       
    linkcolor=red,          
    citecolor=blue,        
    filecolor=magenta,      
    urlcolor=cyan           
} 
\newcommand{\al}{\alpha} 
 
\newcommand{\g}{\gamma}
\newcommand{\de}{\delta} 
\newcommand{\e}{\epsilon}

\newcommand{\La}{\Lambda}

\newcommand{\s}{\sigma}

\newcommand{\w}{\omega}

\newcommand{\pd}{\partial}

\newcommand{\beq}{\begin{equation}}
\newcommand{\eeq}{\end{equation}}
\newcommand{\Beq}{\begin{eqnarray}}
\newcommand{\Eeq}{\end{eqnarray}}
\newcommand{\bml}{\begin{multline}}

\newcommand{\eeqm}{\end{multline}}

\newcommand{\bsp}{\begin{split}}
\newcommand{\esp}{\end{split}}

\renewcommand{\vec}{\overrightarrow}

\renewcommand{\b}[1]{{\bm #1}}

\newcommand{\mc}{\mathcal}

\newcommand{\ra}{\rightarrow}
\newcommand{\req}[1]{Eq.~(\ref{eq:#1})}

\newcommand{\rfig}[1]{Fig.~\ref{fig:#1}}

\newcommand{\nn}{\nonumber}

\DeclareMathOperator{\im}{Im}
\DeclareMathOperator{\re}{Re}

\newcommand{\gam}{0.08}   
\newcommand{\alp}{0.35}   
\newcommand{\ltt}{\de\s_{\rm ltt}}  
\newcommand{\supp}{Supplement} 

\begin{document}

\title{The dynamics of quantum criticality via\\
Quantum Monte Carlo and holography}  
 \author{William Witczak-Krempa}
 \affiliation{Perimeter Institute for Theoretical Physics, Waterloo, Ontario N2L 2Y5, Canada}
\author{Erik S. S{\o}rensen}
 \affiliation{Department of Physics \& Astronomy, McMaster University, Hamilton, Ontario L8S 4M1, Canada}
 \author{Subir Sachdev}
 \affiliation{Department of Physics, Harvard University, Cambridge, Massachusetts, 02138, USA}
 \date{\today}
 \vspace{1.6in}
\begin{abstract} 
Understanding the real time dynamics of quantum systems without quasiparticles constitutes an important
yet challenging problem. We study the superfluid-insulator quantum-critical point of bosons on a two-dimensional 
lattice, a system whose excitations cannot be described in a quasiparticle basis.
We present detailed quantum Monte Carlo results for two separate lattice realizations:
their low-frequency conductivities are found to have the same universal dependence on imaginary frequency and temperature. 
We then use the structure of the real time dynamics of conformal field theories described by the
holographic gauge/gravity duality to 
make progress on the difficult problem of analytically continuing the Monte Carlo data to real time.
Our method yields quantitative and experimentally testable results on the frequency-dependent conductivity
near the quantum critical point, 
and on the spectrum of quasinormal modes in the vicinity of the superfluid-insulator quantum phase transition.
Extensions to other observables and universality classes are discussed. 
\end{abstract}
\maketitle 
\tableofcontents 
\section{Introduction}
\label{sec:intro}
 
The quasiparticle concept is the foundation of our understanding of the dynamics of many-body quantum systems.
It originated in metallic Fermi liquids with electron-like quasiparticles; but it is also useful in more exotic states, such as the fractional quantum Hall states and one-dimensional Luttinger liquids, which have quasiparticle
excitations not simply related to the electron. However, modern materials abound in systems to which
the quasiparticle picture does {\em not\/} apply,\cite{keimer} and developing their theoretical description remains one the most important
challenges in condensed matter physics. Here we develop a quantitative description of the transport properties of a system without
quasiparticles by combining high precision quantum Monte Carlo with recent results from string theory.

\begin{figure}  
\centering
\subfigure[]{\label{fig:QCP} \includegraphics[scale=.38]{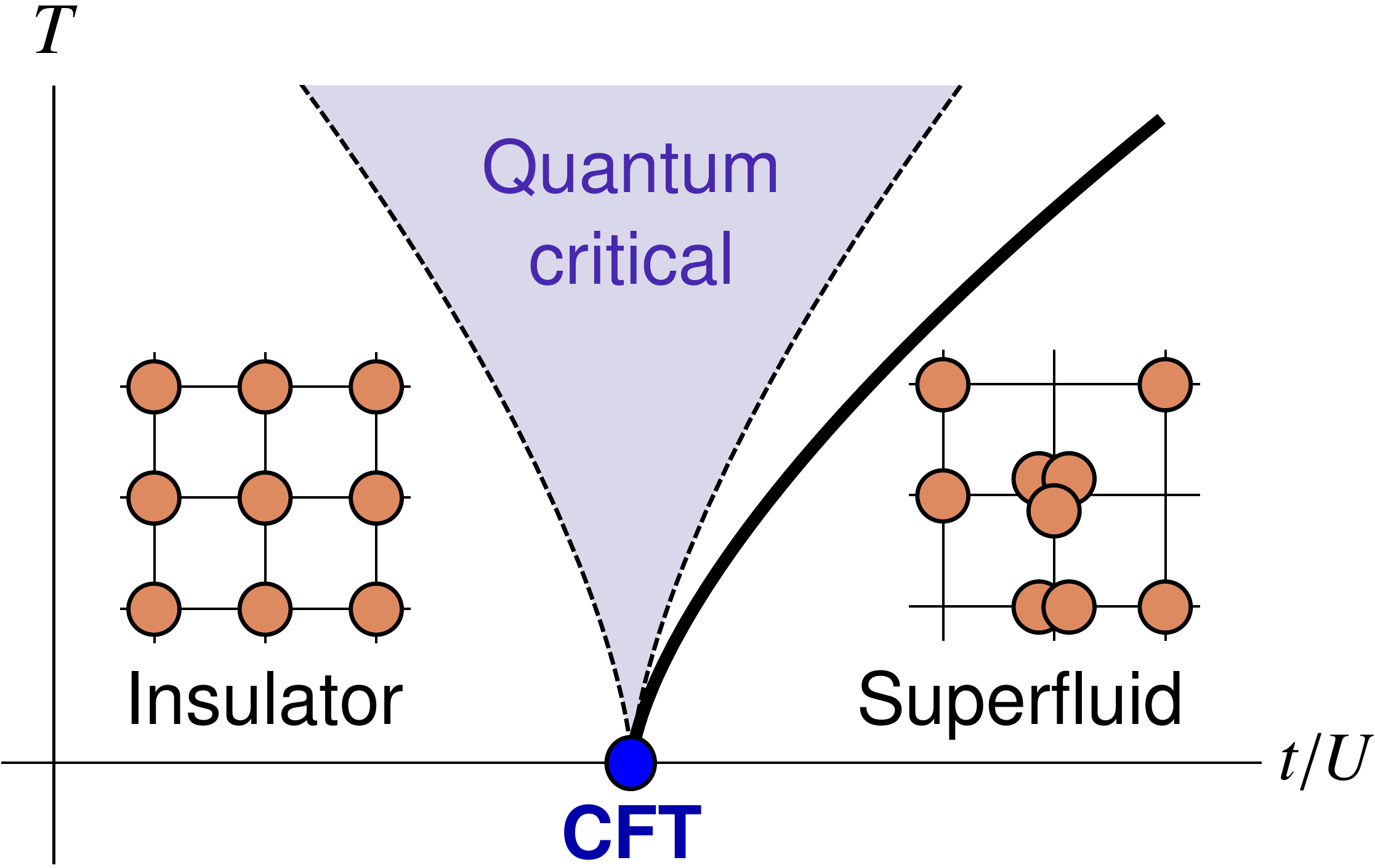}}       
\subfigure[]{\label{fig:sigmaonlyextrap} \includegraphics[scale=.30]{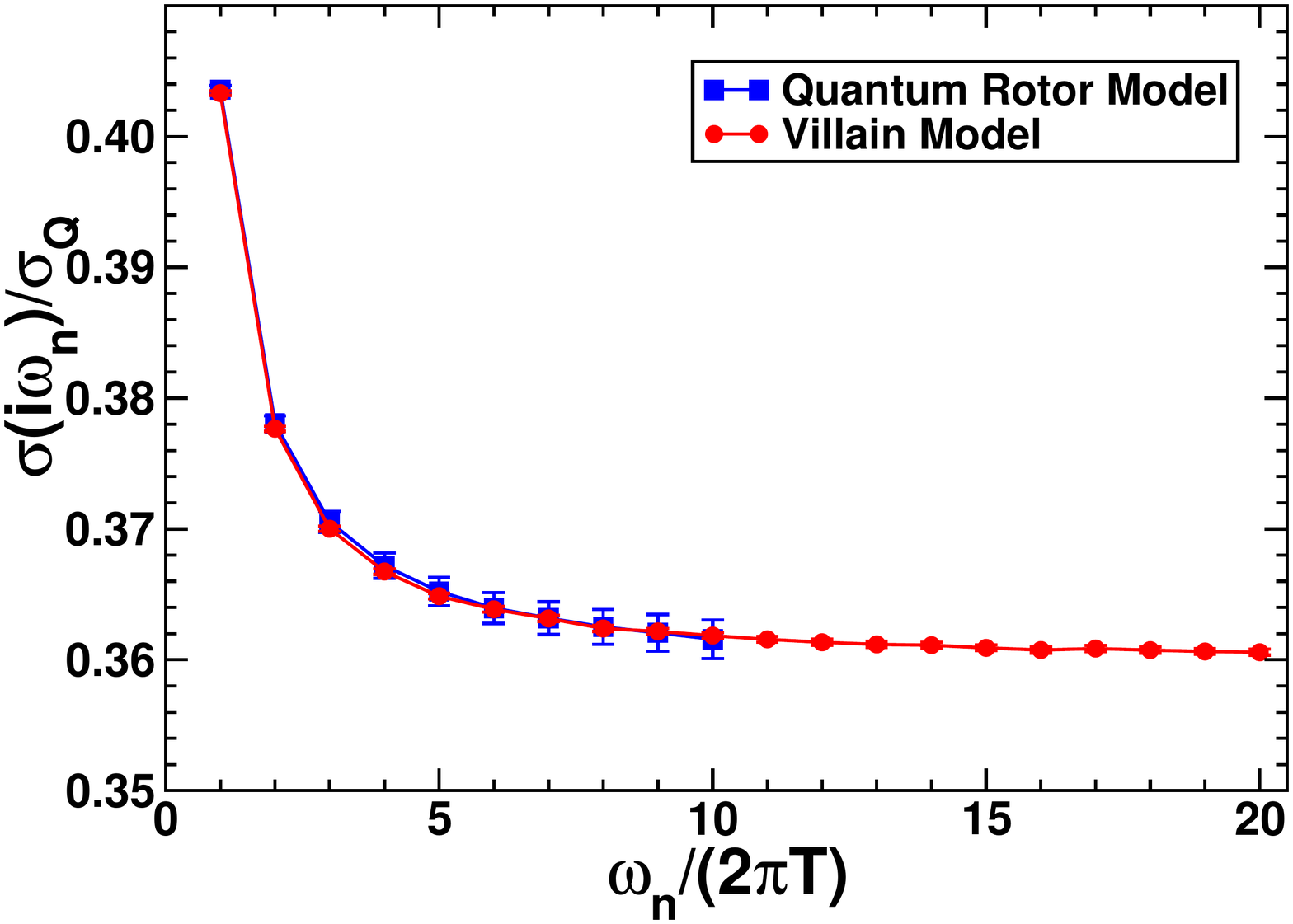}}           
\caption{\label{fig:pd} {\bf Probing quantum critical dynamics} {\bf (a) Phase diagram of the superfluid-insulator quantum phase transition} 
as a function of $t/U$ (hopping amplitude relative to the onsite repulsion) and 
temperature $T$ at integer filling of the
bosons. The conformal QCP at $T=0$
is indicated by a blue disk.  
{\bf (b) Quantum Monte Carlo data} for the frequency-dependent conductivity, $\s$, near 
the QCP along the imaginary frequency axis,
for both the quantum rotor and Villain models. The data 
has been extrapolated to the thermodynamic limit and zero temperature.
The error bars are statistical, and do not include systematic errors arising from the assumed forms of 
the fitting functions, which we estimate to be 5--10\%.   
} 
\end{figure}

We focus on the simplest system without a quasiparticle description: the quantum-critical region of the quantum phase transition
between the superfluid and insulator in the Bose-Hubbard model (BHM) in two spatial dimensions (see \rfig{QCP}).
This quantum critical point (QCP) has special emergent symmetries at low energies, Lorentz and scale invariance,
and the quantum critical dynamics is described by a conformal field theory (CFT). 
We will look at lattice models closely related to the BHM which are more amenable to quantum Monte Carlo (QMC)
studies: the quantum rotor and the Villain models~\cite{villain,Wallin1994}.
This QCP is also of great interest because of its recent experimental realization  
in systems of ultra-cold atoms loaded in optical lattices \cite{spielman,zhang,endres}. 

Our studies are performed in an ``imaginary'' quantum time necessary for efficient simulations.
We obtain high-precision results for thermodynamic observables and for the conductivity along the imaginary
frequency axis at the quantum-critical coupling. 
The results for the conductivity, in units of the quantum of conductance $\sigma_Q=(e^*)^2/h$ (for carriers of charge $e^*$), 
appear in \rfig{sigmaonlyextrap},  
and these are much more precise than earlier studies \cite{sorensen}.
They now convincingly demonstrate that the conductivity, $\sigma$, 
has a non-trivial and universal dependence on $\hbar \omega / k_B T$ \cite{damle}, where $T$ is the absolute temperature. 
Furthermore, the results for the two different lattice realizations agree well with each other, confirming that they are both computing the universal properties of the CFT describing the superfluid-insulator transition. 
Complementary results along the $T=0$ axis appeared recently in Ref.~\onlinecite{gazit}. 

For experimental comparison, we need predictions in {\em real\/} time, and so cannot use the results from \rfig{sigmaonlyextrap} directly. 
Without additional physical input, the analytic continuation from imaginary to real frequencies represents an ill-posed problem in 
which minute errors are invariably magnified by the continuation. We argue here that powerful physical
input can be obtained from a tool that has recently emerged out of string theory, the AdS/CFT or holographic 
correspondence\cite{Maldacena}.
It allows the study of correlated
CFTs (and deformations thereof) without relying on weakly interacting quasiparticles by 
postulating the duality between specific CFTs/string theories. Of special interest is the fact that one can tune the parameters of the CFT such that it remains  
correlated while, simultaneously, the string theory description reduces to classical gravity on Anti de Sitter (AdS) 
spacetime, in one higher spatial dimension. One can thus use well-understood general relativity tools 
to study non-trivial quantum field theories. We will show how such methods allow us to perform the analytic continuation, and
yield much information that is potentially experimentally testable: on the frequency-dependent conductivity and beyond,
and on the positions of poles of response functions in the lower-half of the complex frequency plane, which we have
identified as ``quasinormal modes'' \cite{star2,ws,ws2}.    
 
\section{Simulating Bose-Hubbard quantum criticality} 
\label{sec:qmc}
\begin{figure}
\centering
\subfigure[]{\label{fig:sigmaQR} \includegraphics[scale=.30]{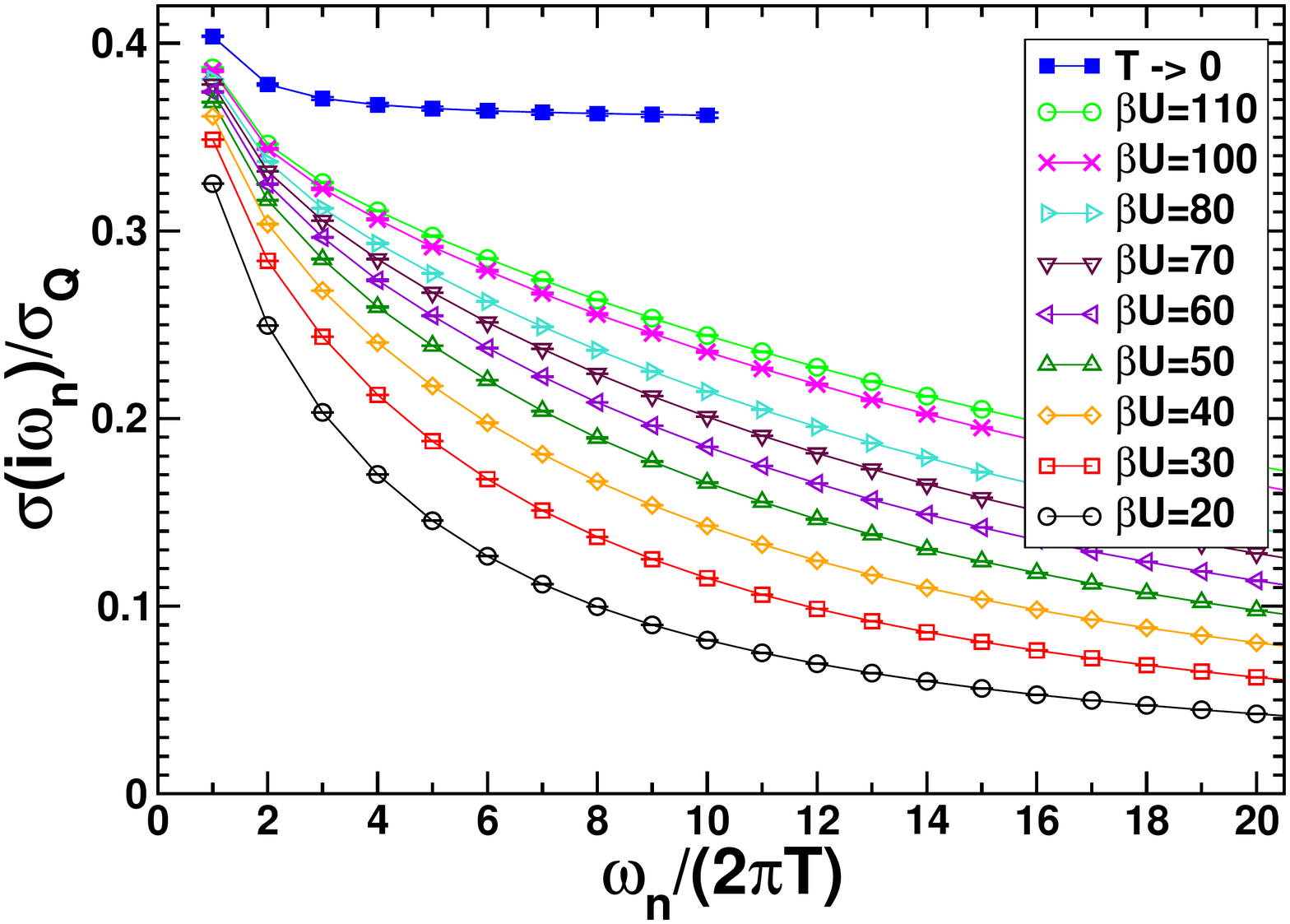}} 
\subfigure[]{\label{fig:sigmaV} \includegraphics[scale=.30]{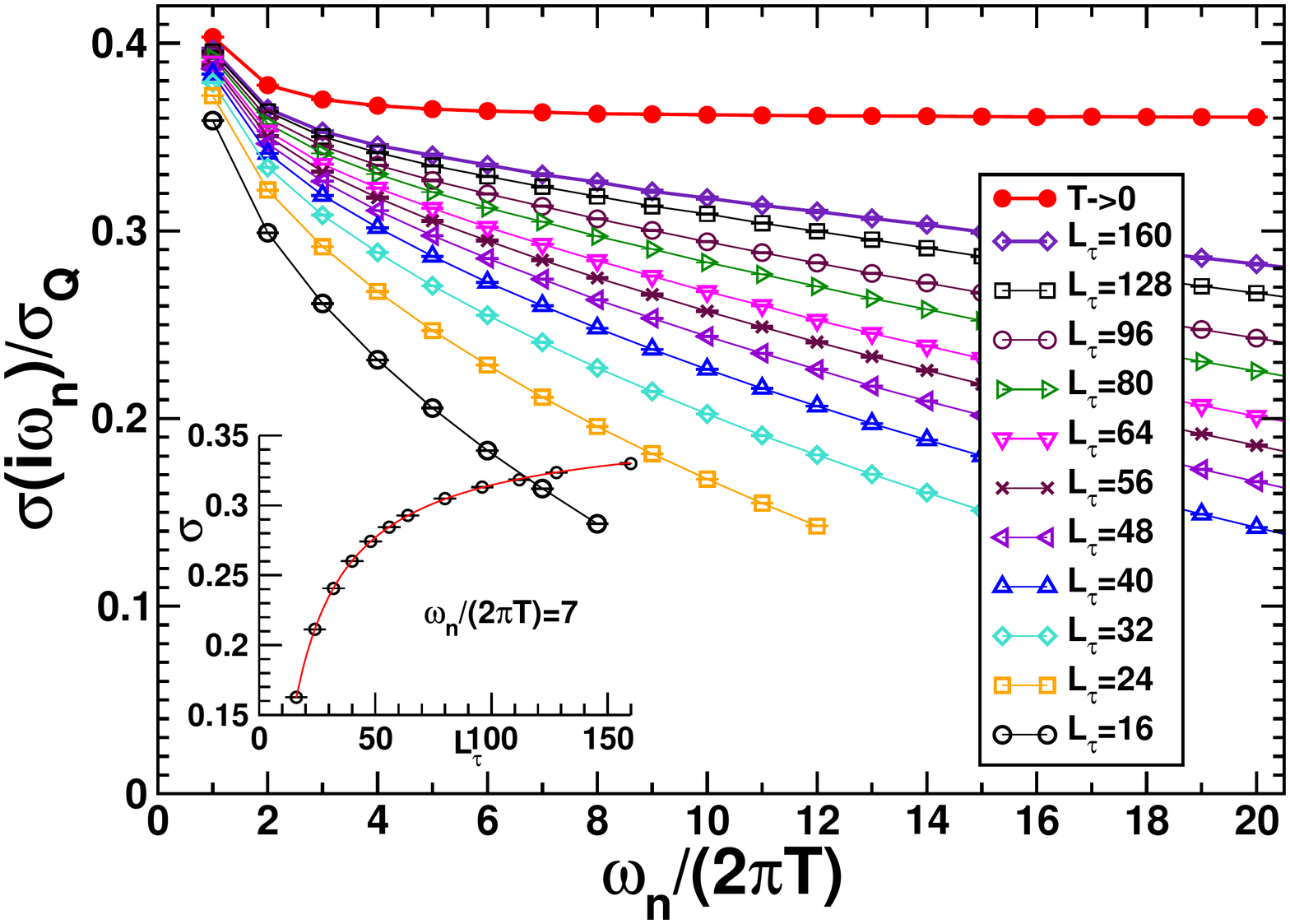}} 
\caption{\label{fig:qmc} {\bf Quantum Monte Carlo data} 
    { (a) Finite-temperature conductivity} for a range of $\beta U$ in the $L\to\infty$ limit for the quantum rotor model at $(t/U)_c$.
The solid blue squares indicate the final $T\to 0$ extrapolated data.
  { (b) Finite-temperature conductivity} in the $L\to\infty$ limit for a range of $L_\tau$ for the  
Villain model at the QCP. 
The solid red circles indicate the final $T\to 0$ extrapolated data. 
The inset illustrates the extrapolation to $T=0$ for $\omega_n/(2\pi T)=7$. The error bars are
statistical for both a) and b). 
} 
\end{figure}   

The extensively studied BHM realizes in a transparent fashion the superfluid-insulator transition of interest to us;
it is defined by the Hamiltonian:  
\begin{align}\label{eq:bhm}
  H=-t\sum_{\langle i,j\rangle}b_i^\dag b_j-\mu\sum_i n_i+\frac{U}{2}\sum_i n_i(n_i-1)\, , 
\end{align}
where $b_i^\dag$ is the creation operator for a boson at site $i$ and $n_i=b_i^\dag b_i$ measures 
the occupation number. 
Tuning $t/U$ at commensurate filling results in a continuous quantum phase transition from a Mott insulator to a superfluid,
as shown in the phase diagram in \rfig{QCP}.  
The intervening conformal QCP is characterized by a U(1) conserved charge  
and belongs to the so-called (2+1)D $XY$ critical universality class. This is the simplest non-trivial CFT in two   
dimensions with a continuous symmetry, and it describes a wide range of critical systems. 
It is strongly correlated so that many of its basic finite-temperature properties remain unknown.  

We performed QMC simulations at $\mu=0$, corresponding to integer filling, on a quantum rotor model, and its Villain version~\cite{sorensen92,Wallin1994}; these are closely related to the BHM  
and have been shown to have QCPs in the same universality class for $\mu=0$.  
The details of the simulations are discussed in the Methods section.
First, we have determined the temperature scaling of various 
thermodynamic quantities in the quantum critical regime, such as the compressibility (charge susceptibility), 
$\chi$, and heat capacity, $c_V$, and 
have confirmed the CFT predictions:
\begin{align}
\label{eq:amplitudes}
  \chi = A_\chi \frac{k_B T}{(\hbar c)^2}\,, \qquad c_V = A_{c_V} \left(\frac{k_BT}{\hbar c}\right)^2\,,  
\end{align}
where $A_\chi=0.339(5)$ for the Villain model, for which the velocity of ``light'', $c$, is known. This result is close
to the large-$N$ field theory estimate of $0.24$\cite{CSY}. In the case of the quantum rotor model, $c$ is not known  
and so a meaningful quantity to give is the dimensionless ratio  
$W=A_{c_V}/A_\chi=c_V/(k_B T\chi)$, which we found to be $6.2(1)$, in excellent agreement with the  
field theory estimate of $6.14$\cite{CSY}. Combining this with the value of $A_\chi$ for the Villain model,
we find $A_{c_V}=2.1$, which lies close to a recent non-perturbative RG result\cite{adam13}, $1.8$;
the field theory estimate\cite{CSY} is $1.5$.   
Exploiting the universality of Eq.~(\ref{eq:amplitudes}) it is now possible to
estimate $c$ for the quantum rotor model. In the simulations $\hbar=1$ as well as the lattice spacing $a=1$; in that case $c$ has
dimensions of energy and it is then natural to estimate $c/U$. By calculating $\beta \chi U^2=A_\chi/(c/U)^2=3.87(3)$ for the quantum rotor
model combined with  the previous result for $A_\chi$ obtained from the Villain model we then find at the critical point $c/U=0.29(1)$ in complete accordance
with a spin-wave estimate yielding $c/U=\sqrt{t/(2U)}=0.295$ at the QCP.
To our knowledge, our simulations are the first to determine these universal coefficients.   

We now turn to the main result, namely the imaginary-frequency conductivity in the quantum critical regime, \rfig{sigmaonlyextrap}.  
It was obtained by first extrapolating the finite-size data to the thermodynamic limit, which was facilitated by the
fact that much larger system system sizes were used than previously. Second, we 
extrapolated to zero temperature to obtain the \emph{universal} scaling dependence, the latter procedure being shown in 
Figs.~\ref{fig:sigmaQR} and \ref{fig:sigmaV}.
Both models, although distinct at the lattice level, show the same conductivity confirming the universality of our results. 
As has been mentioned in the introduction, in order to get the observable real-time conductivity one needs to
perform a difficult analytic continuation. Our main claim is that holography can be of practical help in this, 
and below we describe the crux of the method. 

\section{A hand from String theory} 
\begin{figure}
  \centering
  \includegraphics[scale=.4]{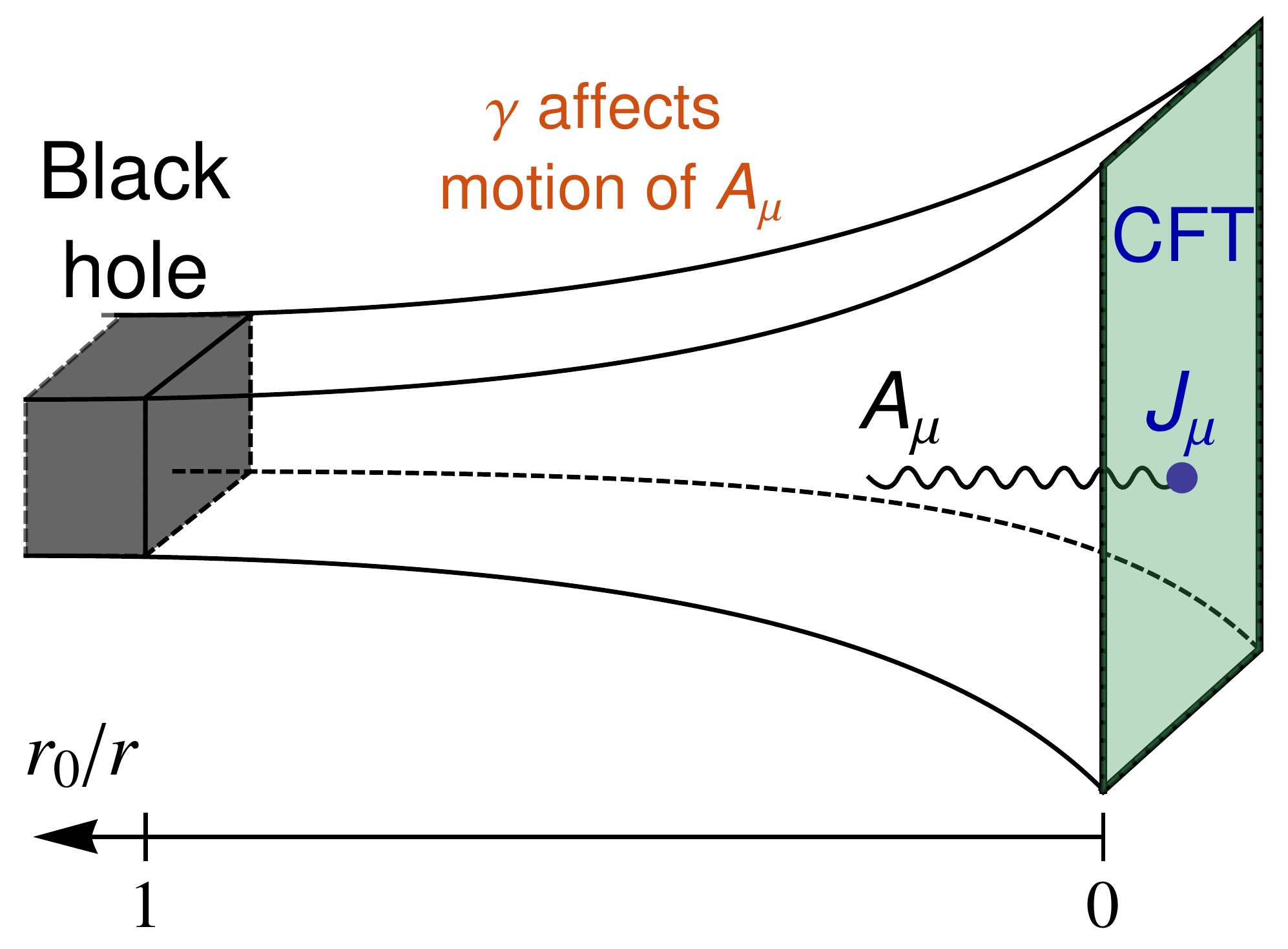}
  \caption{{\bf Holographic spacetime}, which is asymptotically AdS and contains a planar black hole. The current operator of the CFT, $J_\mu$, is
holographically dual to a gauge field, $A_\mu$, in the higher-dimensional bulk. The temperature associated
with the horizon of the black hole is equal to the temperature of the CFT. }  
  \label{fig:ads}
\end{figure} 
 
\begin{figure}[h]
\centering%
\subfigure[]{\label{fig:fit}\includegraphics[scale=.33]{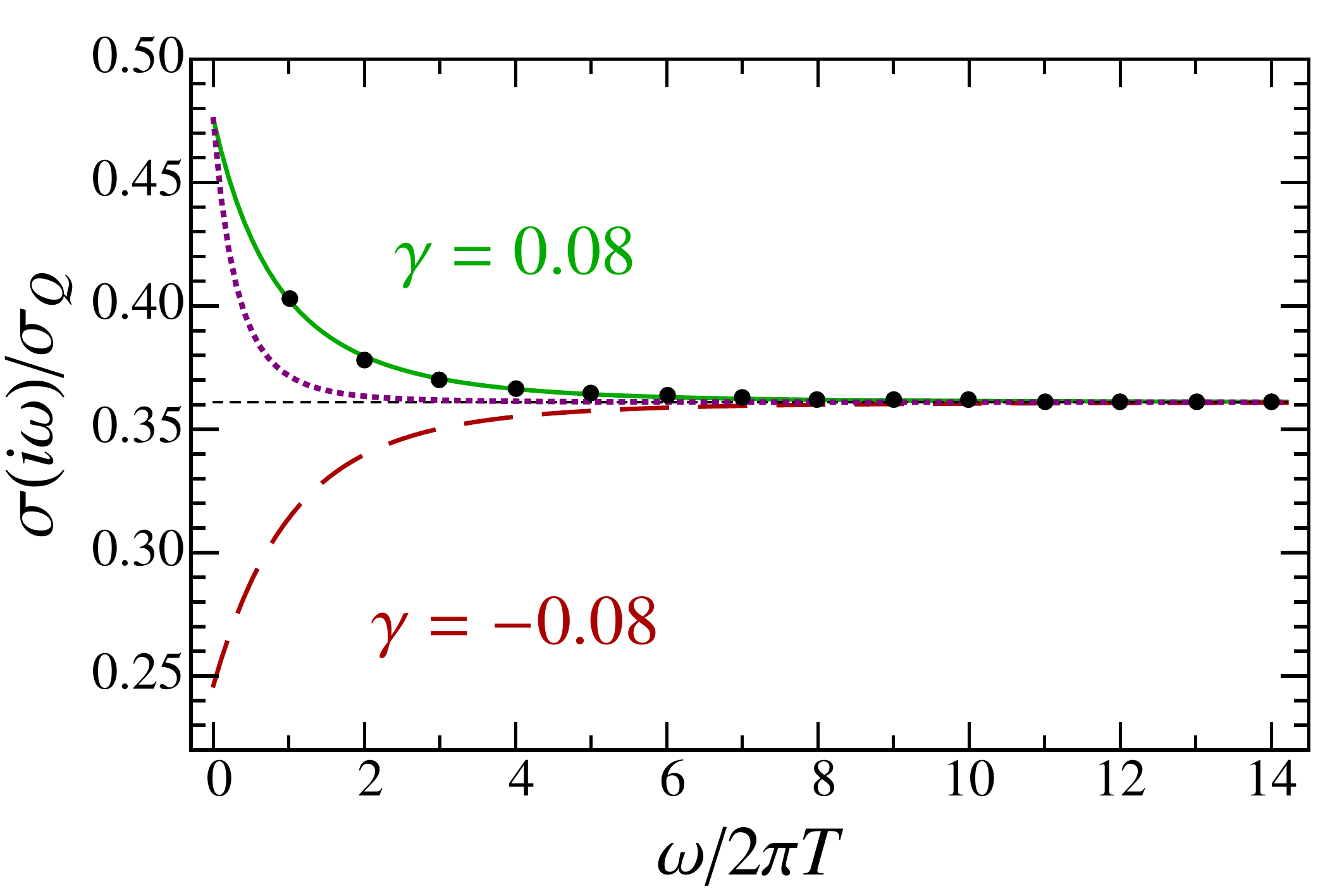}}    
\subfigure[]{\label{fig:cont} \includegraphics[scale=.5]{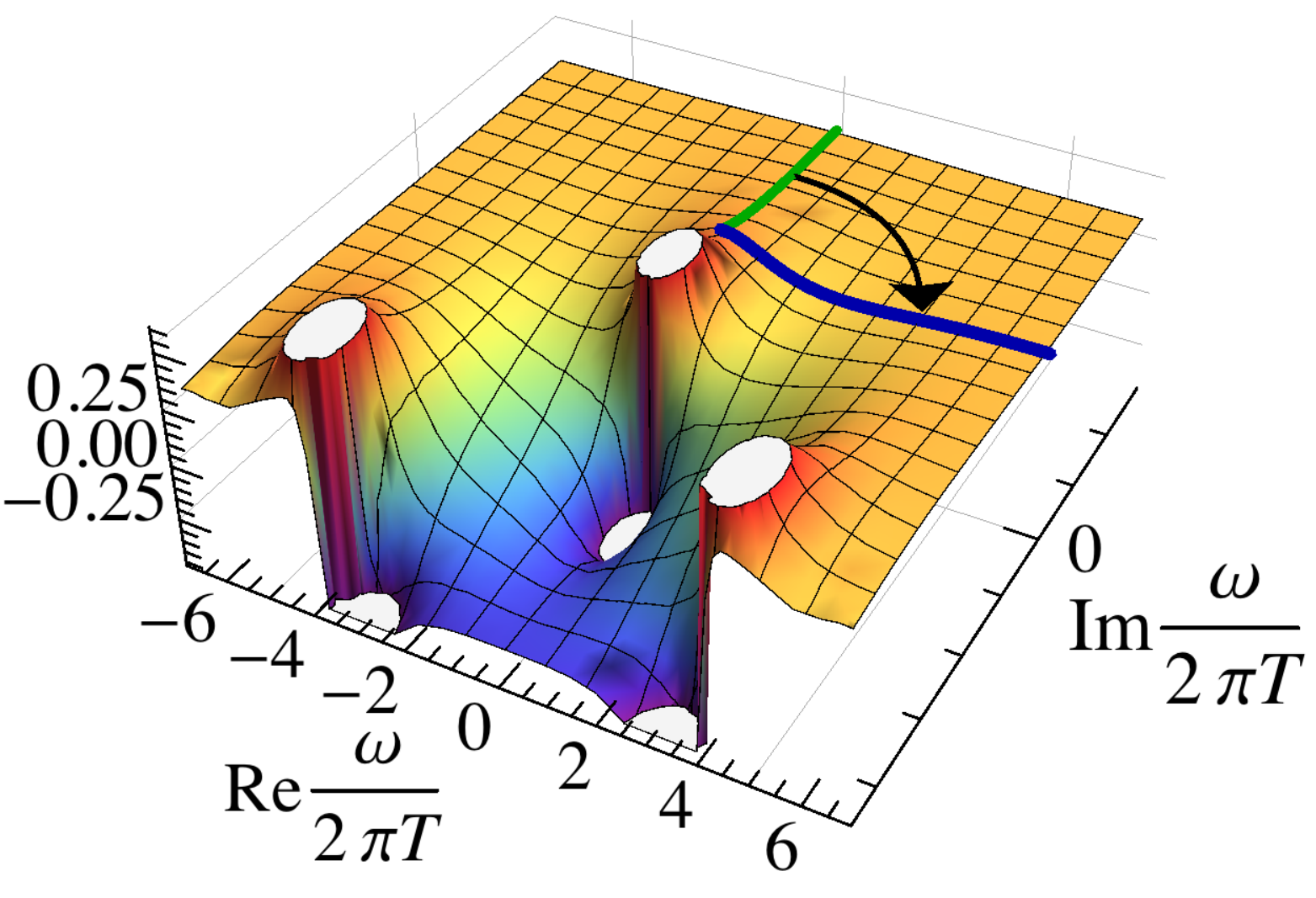}}\\ 
\subfigure[]{\label{fig:cont2} \includegraphics[scale=.33]{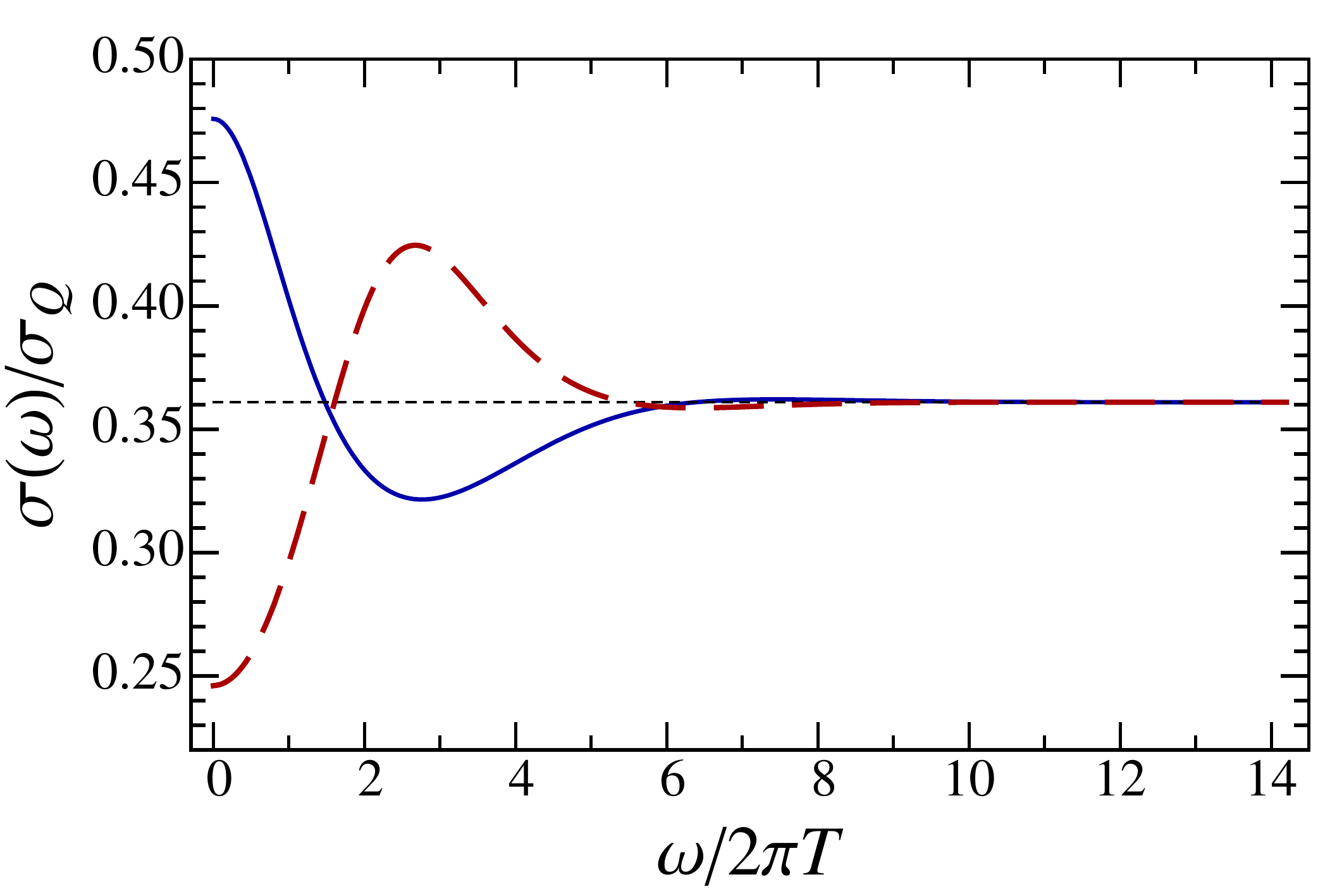}} 
\caption{\label{fig:ads-qmc} {\bf Holographic continuation.} {\bf (a)} The markers represent Monte Carlo
data for the conductivity at the superfluid-insulator QCP at imaginary frequencies, see \rfig{sigmaonlyextrap}. The solid green line 
is the best fit to the holographic conductivity, obtained when $\g=\gam$, with a rescaling of  
the $\w/T$-dependence by $\alpha=\alp$ (the purple dotted line is without the rescaling). 
The red dashed curve is for $\g=-\gam$, and suggests that a vortex-like $\s$ 
does not occur. 
{\bf (b)} Real part of the holographic conductivity evaluated at complex frequencies, where
the imaginary/real axis dependence is highlighted by the green/blue line.  
The arrow represents the continuation procedure. {\bf (c)} Resulting 
conductivity at real frequencies (solid blue line).  
The dashed line is the vortex-like response obtained for $\g=-\gam$.  
}  
\end{figure}

We first briefly summarize the holographic computation of $\s(\w/T)$; we refer the reader
to number of reviews on AdS/CFT aimed towards condensed matter  
applications\cite{sean-rev,mcgreevy-rev,sachdev-rev08,ss-rev,sachdev-rev}, and a brief discussion in the \supp. 
The key ingredient in the calculation     
is that a current operator in the CFT, $J_\mu(t,x,y)$, maps to a dynamical gauge field in the higher dimensional gravitational
theory, $A_\mu(t,x,y;r)$, where $r$ is the coordinate along the extra dimension, see \rfig{ads}. 
The spacetime in which the gauge field evolves is described by the metric:
\begin{align}
  ds^2=\frac{r^2}{L^2}\left[-f(r)dt^2+dx^2+dy^2\right]+\frac{L^2 dr^2}{r^2f(r)} \,, 
\end{align}
where $f(r)=1-r_0^3/r^3$, and $L$ is the radius of AdS$_4$. It asymptotically tends to AdS$_4$ as
$r\ra\infty$, and contains a black hole whose event horizon is located at $r=r_0$.  The latter allows for  
a finite temperature in the boundary CFT, which is in fact determined by the position of the horizon, $r_0=T(4\pi L^2/3)$.  
Heuristically, the Hawking radiation emanating from the black hole escapes 
to $r=\infty$ and ``heats up'' the boundary, where the CFT exists.   
The behavior of the gauge field is determined by extremizing the action\cite{ritz,myers11}:   
\begin{align}\label{eq:S_bulk}
  S_{\rm bulk}=\int d^4x \sqrt{-g}\left(
  -\frac{1}{4g_4^2}F_{ab}F^{ab}+\g \frac{L^2}{g_4^2}C_{abcd}F^{ab}F^{cd}\right)\,,
\end{align}
where $F_{ab}=\pd_a A_b-\pd_b A_a$ is the field strength (roman indices run over 
$t,x,y$ and $r$) while $g_4$ is the bulk gauge coupling, which   
determines the $T=0$ conductivity of the CFT: 
$\s(\w/T\ra\infty)=1/g_4^2$. $C_{abcd}$ is the Weyl tensor, i.e.\ the traceless part of the Riemann curvature tensor, and $\g$
a dimensionless coupling.
The conductivity is obtained  
by solving the modified Maxwell equation associated with \req{S_bulk} for Fourier modes with frequency $\w$.
(The spatial momentum is set to zero.)
We finally use the AdS/CFT relation to obtain the conductivity:   
\begin{align}
  \s(\w/T)=\frac{ir^2}{g_4^2L^2\w}\frac{\pd_r A_y(\w,r)}{A_y(\w,r)}\Big|_{r=\infty}\,.  
\end{align}
When $\g=0$, the conductivity corresponds to that of a supersymmetric Yang-Mills theory with gauge group SU$(N_c)$
in the large-$N_c$ limit\cite{abjm}. In that special limit, the conductivity is frequency independent, $\s(\w/T)=1/g_4^2$,  
due to an emergent electric-magnetic self-duality in the gravitational description\cite{m2cft}.  
The $\g$ term in \req{S_bulk} breaks this self-duality, allowing one  
to probe a wider spectrum of conductivities\cite{myers11}. When $\g>0$ ($<0$) the conductivity has a peak (dip)
near zero frequency, as shown in \rfig{cont2}, thus resembling the conductivity arising from a particle (vortex) description
of the response. 
We refer to these two types of 
responses as particle- and vortex-like, respectively. It is not \emph{a priori} clear which of those two types 
arises in the QCP of the BHM. Below, we settle this question with the help of holography.  
 
\subsection{Holographic continuation}
We now describe the analytic continuation procedure used to extract the observable conductivity.
It is important that the continuation be done on the universal QMC data, resulting from extrapolations to \emph{both}
the thermodynamic and zero-temperature limits in order to avoid contaminating the final result with non-universal dependence. 
Our numerical analysis is the first to adhere to this prescription, originally put forward in Ref.~\onlinecite{damle}.   
Given the imaginary time data, one usually attempts an analytic continuation to real frequencies using standard procedures,
for instance via maximum entropy methods or Pad\'e approximants\cite{smakov}. However, all known continuation procedures are uncontrolled, and in the absence 
of physical input, partly rely on chance. 
We argue that the conductivity computed using holography can be used to perform the continuation 
in a simple and transparent manner, and this provides physical insights that cannot be obtained using any other method. 

The idea behind the ``holographic continuation'' is to fit the QMC data to the holographic conductivity,
evaluated at imaginary frequencies, while allowing for a rescaling of the frequency axis 
(on which we comment below).  
The best fit, shown in \rfig{fit} in green, is excellent and corresponds to    
$\g=\gam$, the same parameter appearing in \req{S_bulk}. It is interesting to note that this value lies within 
the allowed range obtained in holography, $|\g|\leq 1/12\approx 0.083$. 
We also determine $\sigma(\infty)/\sigma_Q = 0.36$, which is in excellent agreement with  
field theory estimates\cite{cha,fazio} and Monte Carlo simulations\cite{cha,sorensen,gazit}.  
The positivity of $\g$ provides strong    
evidence for particle-like response at the superfluid-insulator QCP. We can indeed evaluate the fitted holographic  
conductivity at real frequencies, trivially
realizing the analytic continuation, see \rfig{cont}, and the final result is the solid line in \rfig{cont2}. In contrast, the 
analogous vortex-like conductivity is ruled out by the data: we plot the conductivity corresponding to $\g=-\gam$ along the
imaginary axis in \rfig{fit} with a red dashed line. Its concave shape is clearly at odds with our simulations. 
We claim that the observed convex behavior of $\s$ along the imaginary frequency axis leads to 
a particle-like response, even when analyzed in the presence of higher order derivative terms in the holographic action.
This statement partly relies on the observation that this behavior follows from the presence of
a pole at $\w\propto -iT$ when $\g>0$ (see next subsection). 
This pole will generally dominate the small frequency response as it lies closest to the origin; it leads to a peak
at small and \emph{real} $\w/T$. We emphasize that it is perturbatively stable to higher order derivative corrections.  
A more systematic discussion of these is beyond the scope of the current paper\cite{will-beyond_weyl}.  

A new ingredient is the need to rescale $\w/T$ by $\al=\alp$. This parameter
does not appear naturally in the holographic procedure described above, and might inform us about
the differences between ``small-$N$'' CFTs and those with simple AdS duals.    
Notwithstanding, such a rescaling of the holographic form is benign  
in that it does not alter the essential properties of $\s$, such as the asymptotics, pole/zero structure, 
or sum rules\cite{ws,sum-rules} (see below). So we can view our analytic continuation as a best fit of the imaginary frequency data
to the positions of the poles and zeros of the conductivity in lower-half of the complex plane, while maintaining their relative positions
in the gravity theory.

There are numerous non-trivial merits of the holographic continuation method. For example, the resulting 
conductivity obeys a sum rule\cite{ws,sum-rules}:
$\int_0^\infty d\w [\re \s(\w/T)-\s(\infty)]=0$, that was derived using AdS/CFT but which was conjectured\cite{ws} to
hold in generic CFTs. It was in fact shown to hold\cite{ws} at the conformal QCP of the quantum O$(N)$ rotor model 
(a large-$N$ extension of the one simulated here) in the $N=\infty$ limit, and for free
Dirac fermions.   
Another interesting point arises from the fact  
$\g$ fixes the entire current auto-correlation function, $\langle J_\mu J_\nu\rangle \sim C_{\mu\nu}(\w,\b k)$. 
Thus, extracting $\g$ from the fit, we can predict  
the \emph{momentum dependence}\cite{ws2,m2cft} of the charge and current response using the conductivity data alone. Other
continuation procedures naturally do not give access to such information. These predictions will be tested in 
further work\cite{us-inPrep}.  
 
\subsection{Fingerprint of excitations} 
The holographic continuation procedure in addition gives access to the excitation spectrum of the 
QCP at finite temperature. The holographic conductivity has poles and zeros occurring  
at \emph{complex} frequencies\cite{ws}, specifically in the lower half-plane $\im\w\leq 0$, as required by causality. 
These are the \emph{quasinormal modes} (QNMs) and can be interpreted as substitutes of quasiparticles in a strongly
correlated setting. Interestingly, on the gravitational side of the AdS/CFT duality these correspond to
damped eigenmodes of the black hole\cite{star2}. From the holographic fit to the QMC data,
we can identify the QNMs of $\s$, the first three of which are shown in \rfig{cont}.   
One of them is particularly important: it is located directly on the imaginary axis closest to $\w=0$, and
was called\cite{ws} the D-QNM due to its damped nature (no real part) and formal relation to the Drude conductivity.  
Such a purely imaginary pole was previously found\cite{will-mit} in a large-$N$ extension of the 
quantum rotor model studied in this work from O(2) symmetry to O$(N)$, as well as in the study of graphene
in the presence of Coulomb interaction\cite{fritz08}.     
It allows a sharp distinction between particle- and vortex-like responses: in the former case the D-QNM is a pole
while in the latter it is zero. 
In addition to that pole, the spectrum contains two infinite branches of QNMs composed of alternating poles and zeros.
\begin{figure} 
\centering
\includegraphics[scale=.4]{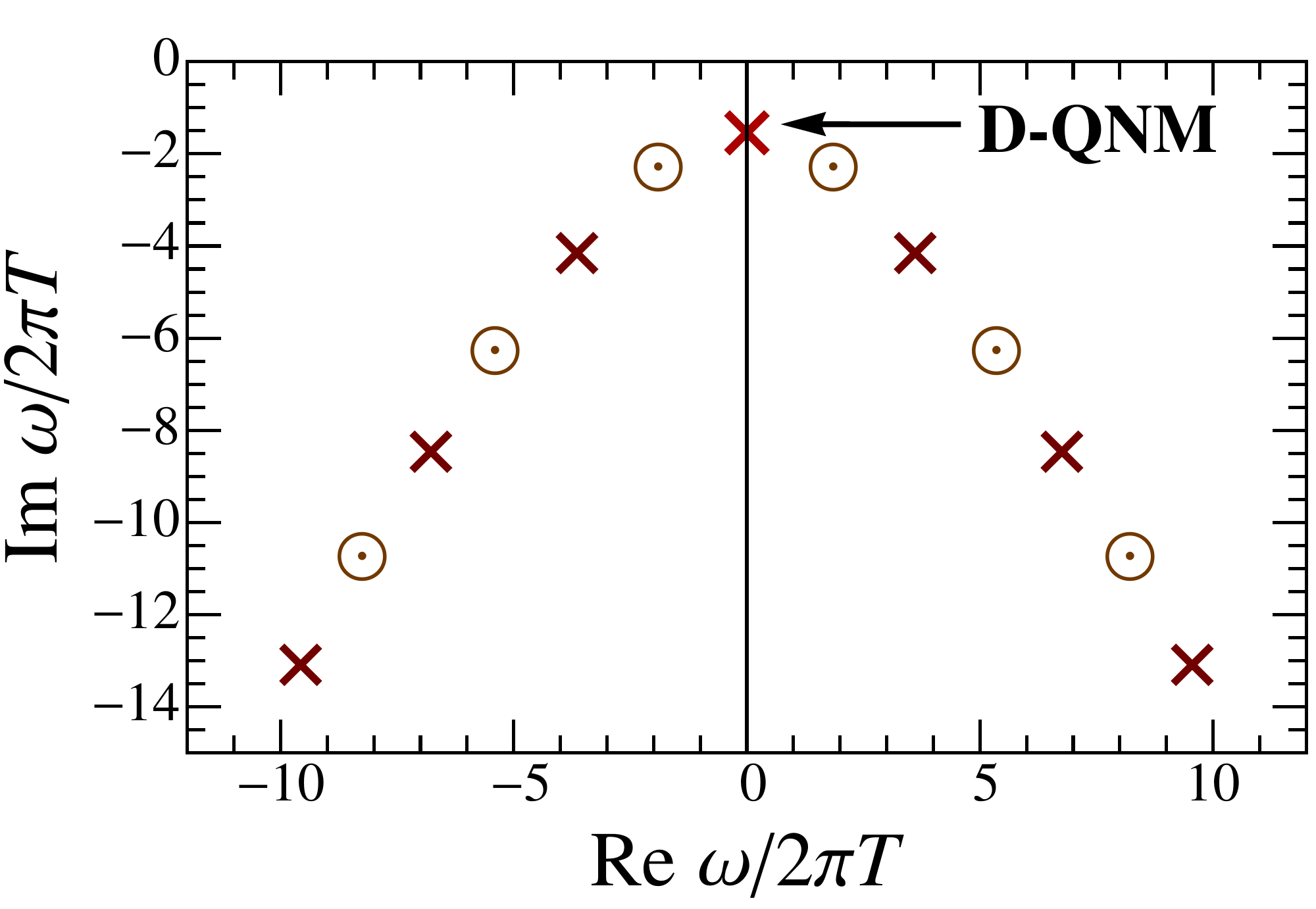} 
\caption{\label{fig:qnm} {\bf Spectrum of quasinormal charge excitations of the superfluid-insulator QCP}. 
The crosses/circles identify poles/zeros of $\s(\w/T)$ in the \emph{complex frequency} plane. The dominant
QNM, labeled D-QNM, is found to be a pole. It gives rise to a peak in $\re\s(\w/T)$ at small frequencies.  
}  
\end{figure}   
 
Insights into the QNM spectrum help to understand why Pad\'e approximants are ill-suited for the analytic  
continuation. In the latter, one matches the power series for  the imaginary frequency conductivity data to that of a rational function of $\w$:
$P_m(\w)/Q_n(\w)$, where $P_m,Q_n$ are polynomials with undetermined coefficients of order $m,n$, respectively. 
Since the number of data points is small, one is limited to relatively small $m,n$. The $m,n$ zeros of $P_m,Q_n$, respectively, 
can be seen as crude approximations to the QNMs described above. A major problem with such an approach   
is that we expect $\s$ to have an infinite number of QNMs, as suggested by the holographic analysis, making it   
impossible for the Pad\'e approximant to capture the entire $\w$-dependence.  
Further, the Pad\'e approximant will generically have spurious poles/zeros, sometimes even in the upper half-plane
$\im\w >0$, violating the causality requirement, and generally the sum rule. 
In contrast, the holographic conductivity avoids such problems, allowing one to tune an infinite sequence of
physically motivated QNMs using a small number of parameters.

Before closing, we note an interesting feature of transport in two dimensions: arguments from hydrodynamics applied to the 
classical regime $\w\ll T$ of the CFT, show that thermal \emph{fluctuations} lead to a weak and universal 
logarithmic divergence of the 
conductivity $\s(\w/T)\sim \ln(T/|\w|)$, such that the d.c.\ conductivity is infinite.
This phenomenon usually goes under the name of \emph{long-time tails}\cite{pomeau-rev,sachdev-rev08,kovtun-rev}. 
It is difficult to derive starting from the original quantum Hamiltonian:
indeed, QMC and conventional quantum Boltzmann approaches, just like leading order holography, do not 
see such an effect. In the case of QMC, this is because data is only available for $|\w|\geq 2\pi T$.
Notwithstanding, we do not expect this small-frequency, classical feature to significantly modify the frequency 
dependence of $\s$ beyond $|\w|\ll T$ and the associated QNM spectrum. This claim is further justified by our estimates in the \supp.

\section{Discussion} 
\label{sec:discuss}

We have shown how the AdS/CFT correspondence can be used in a concrete fashion in conjunction with   
high precision QMC simulations to shed light on the quantum dynamics of correlated  
QCPs. We have illustrated our point by examining the charge conductivity of the conformal QCP of the Bose-Hubbard model
in two spatial dimensions. A holographically derived conductivity was employed to perform the analytic continuation of 
large-scale QMC data, allowing us to obtain the universal scaling function for the conductivity $\s(\w/T)$. 
In doing so, we established the 
particle-like nature of the charge response, as well as the spectrum of (quasinormal) charge excitations, 
which play the role of ``quasiparticles'' in strongly correlated systems.

A validation of our procedure was that the fitting parameter $\gamma$ was 
within the range allowed by holography\cite{myers11} (see also the \supp).  
We also note that we rescaled the $\omega/T$ axis of the holographic theory
by $\al=\alp$ to fit the data: this suggests that holographic 
description of CFTs with small symmetry groups
require significant quantum renormalizations of the $T$ scale of the dynamics, relative to the 
$T$ scale determining thermodynamics.

We emphasize that our results are experimentally relevant, for example for the Mott transition of bosonic 
cold atoms in a optical lattice, which has been recently realized\cite{spielman,zhang,endres}. Measurements of the optical conductivity near 
the QCP could be made in such systems in the near future and confronted with our predictions. 
Although we have restricted ourselves to the analysis of the conductivity, 
one can consider other correlation functions such as  
those involving the complex order parameter field. 
Further, it would be of great interest
to examine other models, such as the 
conformal QCP of the O$(N)$ model for $N>2$, with the case $N=3$ being of particular physical relevance, and see how
the results compare with the $N=2$ case discussed here. More generally, our work has initiated a quantitative confrontation between
holographic theories and realistic condensed matter systems, and has wide scope for extensions. 

\section*{Methods}
 
\begin{figure}  
\centering
\subfigure[]{\label{fig:sigma80} \includegraphics[scale=.30]{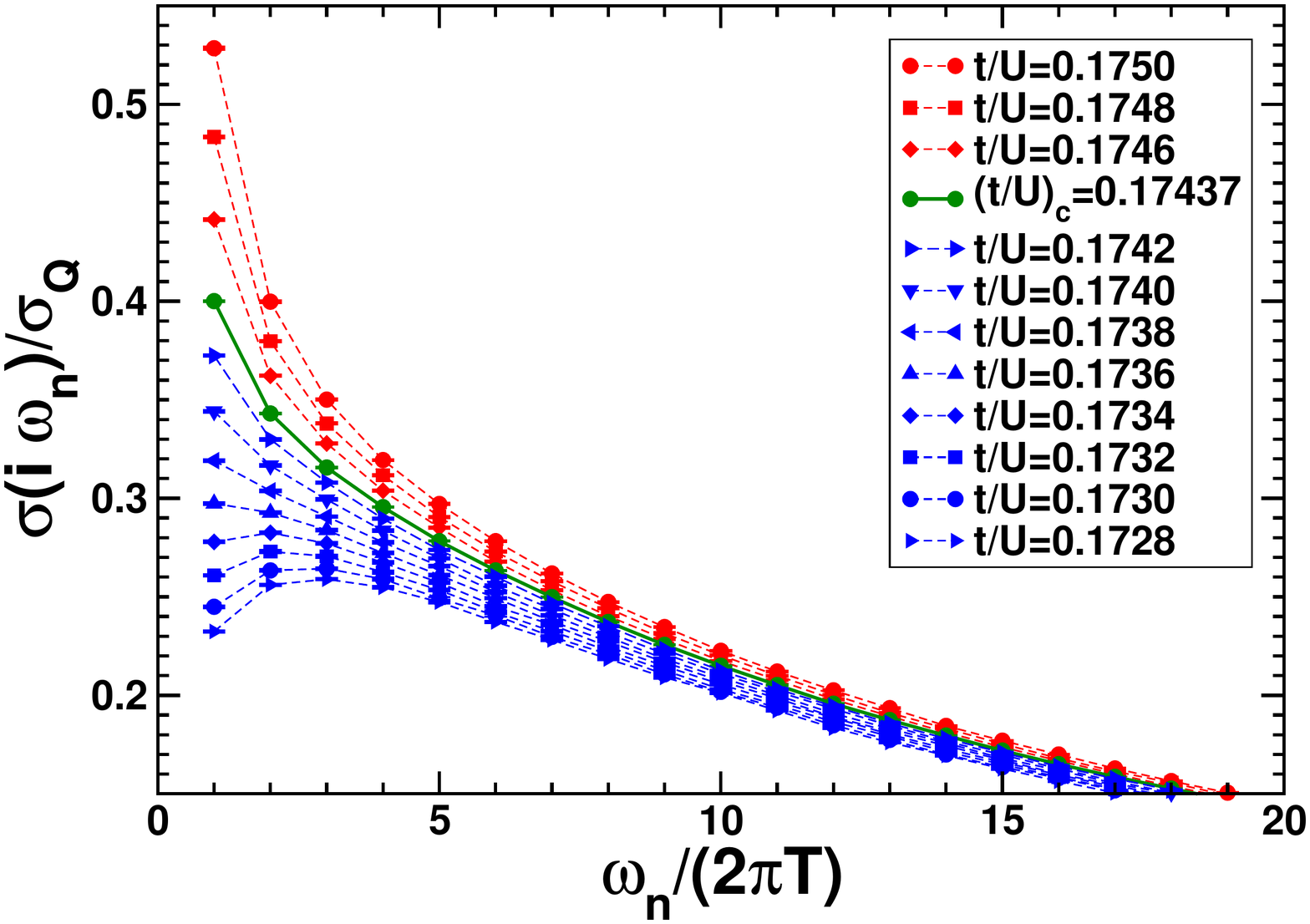}} 
\subfigure[]{\label{fig:crossing} \includegraphics[scale=.30]{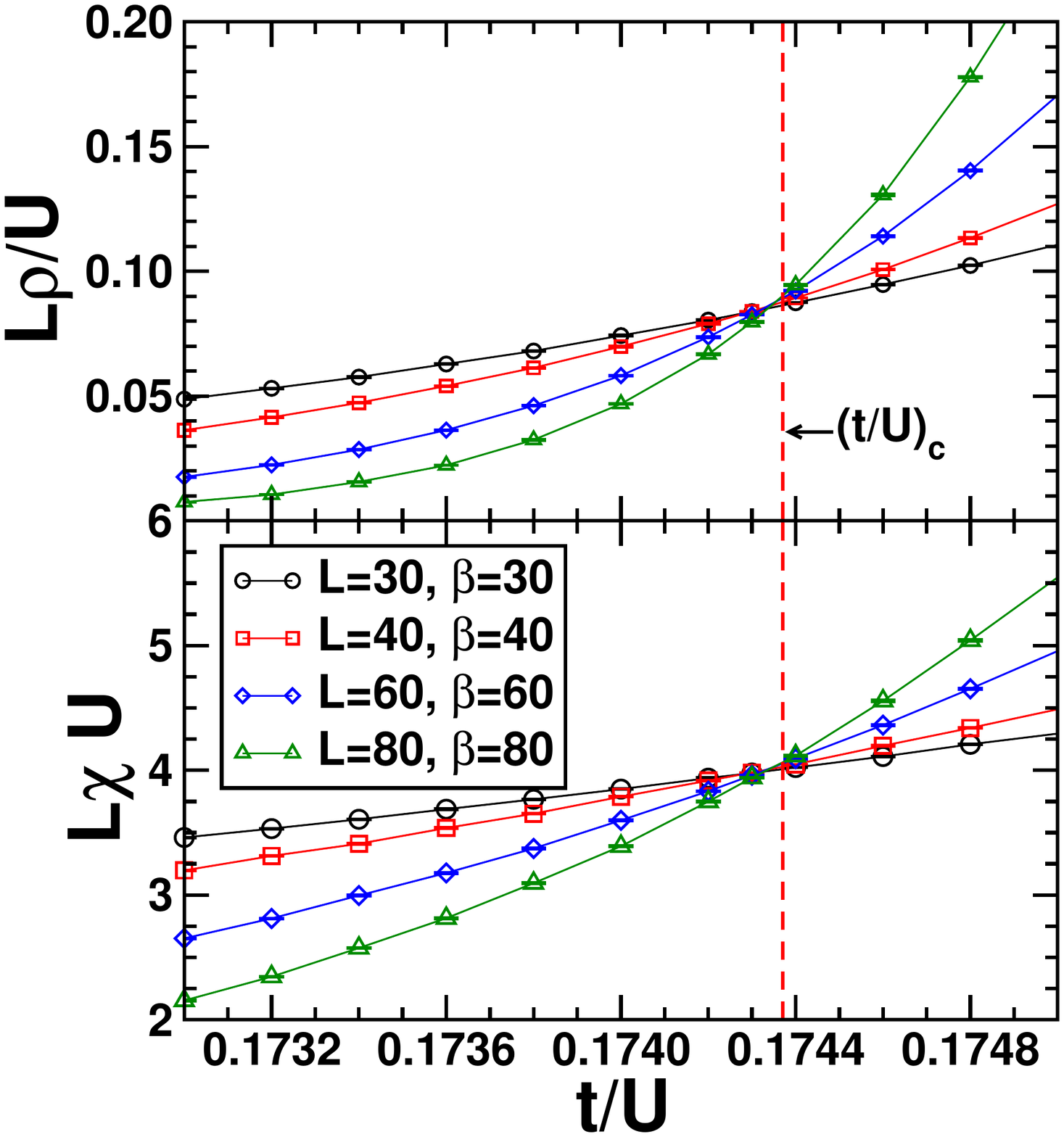}} \\
\caption{\label{fig:num_qcp}  
{\bf (a) Finite-temperature conductivity} for the quantum rotor model at imaginary frequencies in the insulating (blue), quantum critical (green)
and superfluid (red) regimes. Results are for a $80\times 80$ lattice at $\beta U=80$ for a range of $t/U$.
  {\bf (b) Finite-size scaling of the phase stiffness}, $\rho$, and compressibility, $\chi$, to locate the QCP,  $(t/U)_c=0.17437(1)$ (dashed red line) for the quantum rotor model.
Results are shown with $\beta/L=1$.
} 
\end{figure}

The simulations of the quantum rotor model are performed on $L\times L$ lattices at a dimensionless inverse temperature $\beta U$
with a fixed small discretization of the temporal axis $\Delta\tau=0.1$.
We have checked that our results are not affected by the finite $\Delta\tau$.
In contrast, the simulations for the Villain model are performed on $L\times L\times L_\tau$ lattices with $L_\tau$ playing the role of $\beta U$ 
and with an effective hopping strength called, $K$, taking the place of $t/U$.
Typically more than $10^9$ Monte Carlo steps are performed for each simulation.
Both of these models allow for very efficient {\it directed} Monte Carlo sampling~\cite{aleta,aletb}, and as a result, 
we can study systems with up to $320\times 320$ sites with $L_\tau=160$ (Villain model) 
and $110\times 110$ with $\beta U=110,\ \Delta\tau=0.1$ (quantum rotor model).  
As an illustration of how the QCP is detected in the simulations \rfig{sigma80} shows data for the quantum rotor model with a  $80\times 80$ lattice at a fixed $\beta U=80$
as the QCP is traversed by varying $t/U$ from the insulating (blue) through the quantum critical (green) to the superfluid (red) regimes.
The data clearly shows the characteristic ``separatrix'' often seen in experiments on the superconductor to insulator transition in two dimensions~\cite{Haviland1989}. 

While the location of the QCP is known~\cite{sorensen92,aleta,neuhaus,pollet} for the Villain model, $K_c=0.3330671(5)$, 
we have determined the location of the QCP with excellent accuracy for the quantum rotor model as is shown in \rfig{crossing}.
Finite-size scaling predicts that the stiffness $\rho$ and compressibility $\chi$ should scale as $1/L$ at the QCP if $\beta/L$ is kept fixed.
Plots of $L\rho$ and $L\chi$ should then show a clear crossing at the QCP as clearly evident in \rfig{crossing}.
A more refined analysis including corrections to scaling determines our final estimate of the location of the QCP, $(t/U)_c=0.17437(1)$ 
(see \supp).

\section*{Acknowledgments} 
We are grateful to R.~Myers and S.-S.~Lee for answering our questions 
and making many useful suggestions. We further acknowledge insightful discussions with J.~Bhaseen, 
J.~Carrasquilla, K.~Chen, D.~Chowdhury,
A.G.~Green, J.~McGreevy, L.~Pollet, N.~Prokof'ev, S.~Raju, A.~Singh and J.~Sonner.
While the present paper was being completed we learned of Ref.~\onlinecite{pollet}, 
which we have found helpful to improve the accuracy of some of our numerical results. 
E.S.S.\ acknowledges allocation of computing time at the Shared Hierarchical Academic Research Computing Network (SHARCNET:www.sharcnet.ca)
and support by NSERC. S.S.\ was supported by the NSF under Grant DMR-1103860 and by the Templeton Foundation.
This research was supported in part by Perimeter Institute for Theoretical Physics (W.W.-K.\ and S.S.). 
Research at Perimeter Institute is supported by the Government of Canada through Industry Canada 
and by the Province of Ontario through the Ministry of Research and Innovation.   \\
 
\noindent {\bf Author contributions}\\
E.S.S.\ performed the large scale simulations. All authors contributed equally to the
analysis of the data and writing of the manuscript.\\  
{\bf Additional information} \\
The authors declare no competing financial interests. Supplementary information accompanies this paper 
on www.nature.com/naturephysics. Reprints and permissions information is available online at 
http://npg.nature.com/reprintsandpermissions. Correspondence and requests for materials should be addressed to W.W.-K.

\appendix 

\section{Quantum Monte Carlo} 
\label{sec:app-qmc}
Our starting point for the QMC calculations is the quantum rotor model defined in terms of phases $\theta_{\bf r}$ living on the sites, $\bf r$, of a two-dimensional square lattice:
\begin{equation}
H_{\text{qr}}=\frac{U}{2}\sum_{\bf r} 
\frac{1}{2}\left( \frac{1}{i}\frac{\partial}{\partial
    \theta_{\bf r}} \right)^2-\mu\sum_{\bf r}\frac{1}{i}\frac{\partial}{\partial
        \theta_{\bf r}}
-\sum_{\langle {\bf r},{\bf r'}\rangle }
t \cos(\theta_{\bf r}-\theta_{\bf r'}) \ .
\label{hqr}
\end{equation}
Here $\frac{1}{i}\frac{\partial}{\partial \theta_{\bf r}}$ 
is usually identified with the angular momentum of the quantum rotor at site ${\bf r}$, which is the canonical conjugate 
of $\theta_{\bf r}$,  but it can also be viewed as the deviation from an average (integer) particle 
number and this model is therefore in the same universality class as the Bose Hubbard model, Eq.~(1) in the main text. $U$ then plays the role of the on-site repulsive interactions
hindering large deviations from the mean particle number and $t$ is the hopping between nearest neighbor sites. Finally, we include a chemical potential $\mu$ and we see
that $\mu=0$ corresponds to the case of integer filling, the case we focus on here.

If a standard Trotter decomposition where the imaginary time, $\hbar\beta$, is divided into $M$ time slices of size $\Delta\tau=\beta/M$ is performed, it can be shown that the partition
function can be written in terms of an {\it integer-valued} current ${\bf J}=(J^x,J^y,J^\tau)$ with $J^\tau$ the angular momentum (or particle number) in the following manner:
\begin{align}
Z_{QR}\approx 
{\sum_{\{\bf J\}}}'
\exp\left\{-
\sum_{({\bf r}, \tau)}\left[
\Delta\tau U\left( \frac{1}{2}\left[J^\tau_{({\bf r},\tau)}\right]^2-\frac{\mu}{U} J^\tau_{({\bf r},\tau)}\right)
-\ln\left(I_{J^x_{({\bf r},\tau)}}(t\Delta\tau)\right)
-\ln\left(I_{J^y_{({\bf r},\tau)}}(t\Delta\tau)\right)
\right]\right\} \,,
\label{eq:ZQR}
\end{align} 
with $I_J$ the modified Bessel function of the first kind, of order $J$. 
Here the ${\sum}'$ denotes the fact that the summation over ${\bf J}$ is constrained to divergence-less configurations making the
summation over the integer valued currents highly non-trivial to perform. Fortunately, this problem 
can be resolved using advanced {\it directed} Monte Carlo techniques~\cite{aleta,aletb}.

The closely related Villain model arises by approximating the $\cos\theta$ term by a sum of
periodic Gaussians centered at $2\pi m$:
$
\exp({t\Delta\tau\cos(\theta)})\simeq \exp({t\Delta\tau})\sum_m\exp({-\frac{1}{2}t\Delta\tau
(\theta - 2\pi m)^2})
$,
preserving the periodicity of the Hamiltonian in $\theta$. 
Employing this simplification leads to the Villain model:~\cite{sorensen92,Wallin1994} 
\begin{equation}
Z_{V}\approx 
{\sum_{\{\bf J\}}}'
\exp\left[-\frac{1}{K}
\sum_{({\bf r}, \tau)}\left(
\frac{1}{2}{\bf J}^2_{({\bf r},\tau)}-\frac{\mu}{U} J^\tau_{({\bf r},\tau)}  
\right)\right] \ . 
\label{eq:ZV}
\end{equation}
Here $L_\tau$ takes the place of the dimensionless inverse temperature $\beta U$ and varying $K$ is analogous to varying $t/U$ in the
quantum rotor model. Also in this case is the summation over ${\bf J}$ constrained to divergence-less configurations.

For both models it turns out that the frequency dependent conductivity in units of the quantum of conductance $\sigma_Q=(e^*)^2/h$ (for carriers of charge $e^*$) can be calculated by evaluating
\begin{equation}
\sigma(i\omega_n)/\sigma_Q =
\frac{1}{L^{d-2} k}\left< \left|\frac{1}{L}\sum_{({\bf r},\tau)}e^{i\omega_n\hbar\tau} J^x_{({\bf r},\tau)}\right|^2\right>,
\end{equation}
which is dimensionless in $d=2$. Here $n$ is an integer labeling the Matsubara frequency $\omega_n$. 

As mentioned above, we can interpret $J^\tau$ as the particle number. We then find for the total particle number, $N$: 
\begin{equation}
N=\frac{1}{M}\sum_{({\bf r},\tau)}J^\tau_{({\bf r},\tau)}\,,
\end{equation}
where $M$ is the number of (imaginary) time slices.   
With $\left<n\right>=\left<N\right>/L^d$, the particle number per site, it follows that the compressibility 
for the quantum rotor model, $\chi$, is given by
\begin{equation}
\chi=\frac{\partial\left<n\right>}{\partial \mu}=\frac{\beta}{L^d}\left(\left<N^2\right>-\left<N \right>^2\right)\ .
\label{eq:kappaQR}
\end{equation}
Similarly, the stiffness for the quantum rotor model is given by
\begin{equation}
\rho=\frac{1}{L^{d-2}\hbar\beta }\left< \left(\frac{1}{L}\sum_{({\bf r},\tau)}J^x_{({\bf r},\tau)}\right)^2\right>,
\label{eq:rhoQR}
\end{equation}
We note that, in the simulations the temperature is measured in units of $U$, so that one typically evaluates $\chi U$ and $\rho/U$.
Analogous expressions for $\rho$ and $\chi$ are well known for the Villain model~\cite{Wallin1994}. We also point out that all simulations
reported here for both the Villain and quantum rotor model have been performed with $\mu=0$.
Both the quantum rotor and Villain models belong to the (2+1)D $XY$ critical universality class. This can be confirmed by calculating
the correlation length exponent, $\nu$, defined through the divergence of the correlation length $\xi\sim\delta^{-\nu}$ where
$\delta$ is the distance to the QCP. High precision estimates for the Villain model~\cite{aleta} have found $\nu=0.670(3)$ and,
by a direct evaluation of $L\rho'$ at $(t/U)_c$, we here estimate it to be $\nu=0.678(8)$ 
for the quantum rotor model
in excellent agreement. 
Both results are in good accordance with recent high-precision estimates~\cite{campostrini} for the 3D $XY$ model.

\begin{figure}
\centering
\subfigure[]{\label{fig:Critpoint} \includegraphics[scale=.30]{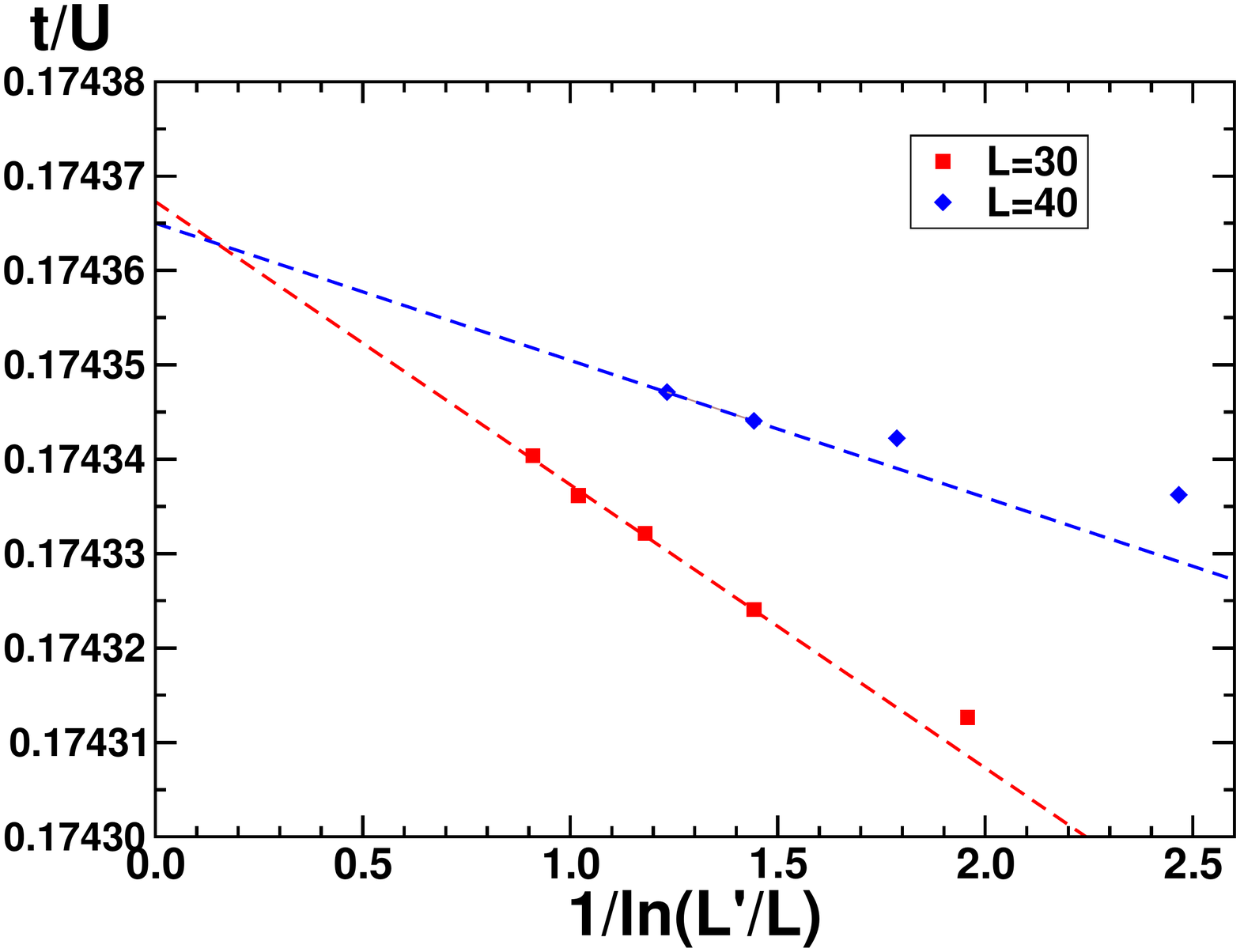}} 
\subfigure[]{\label{fig:BKappaQR} \includegraphics[scale=.30]{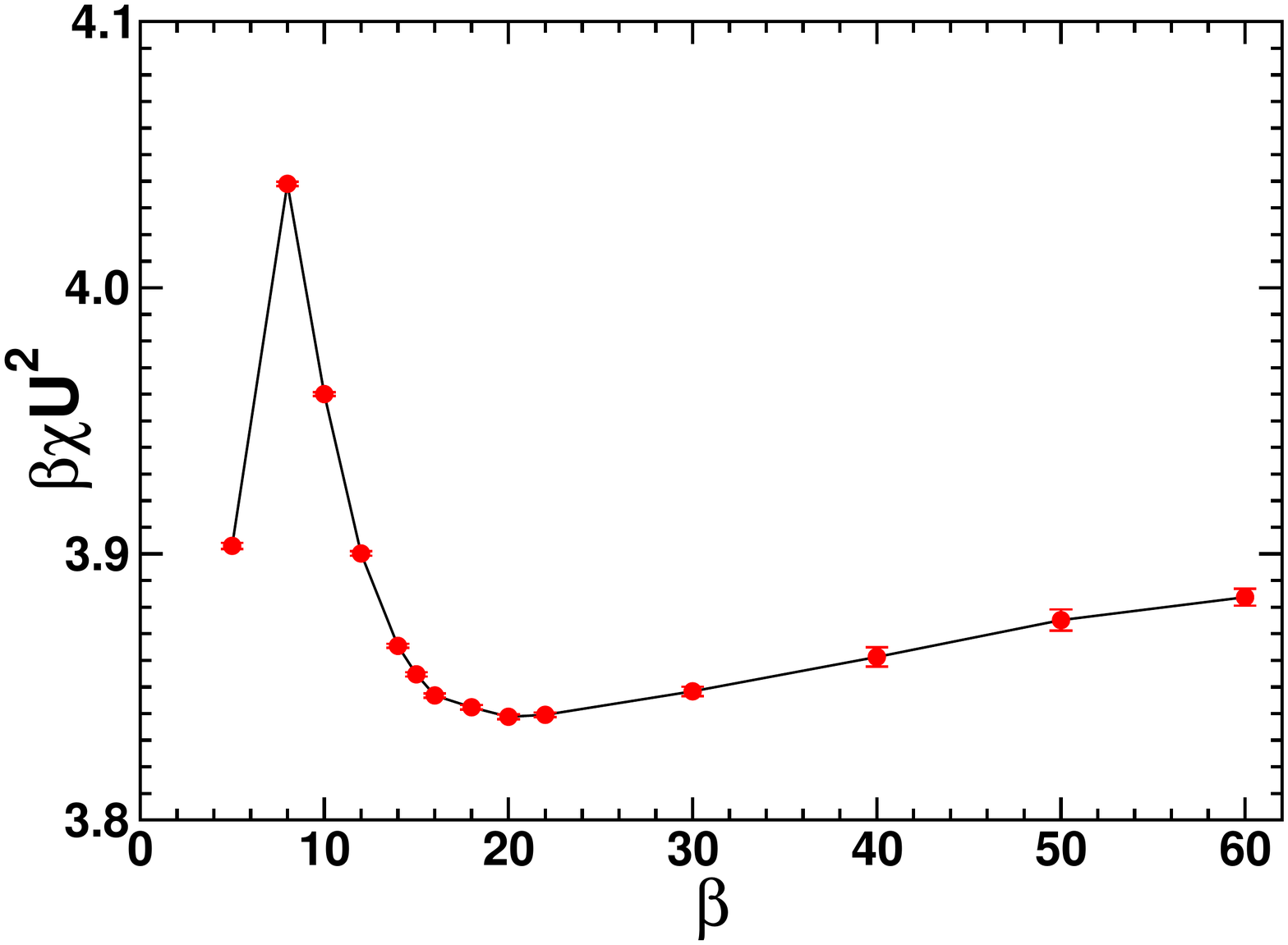}} \\
\subfigure[]{\label{fig:WQR} \includegraphics[scale=.30]{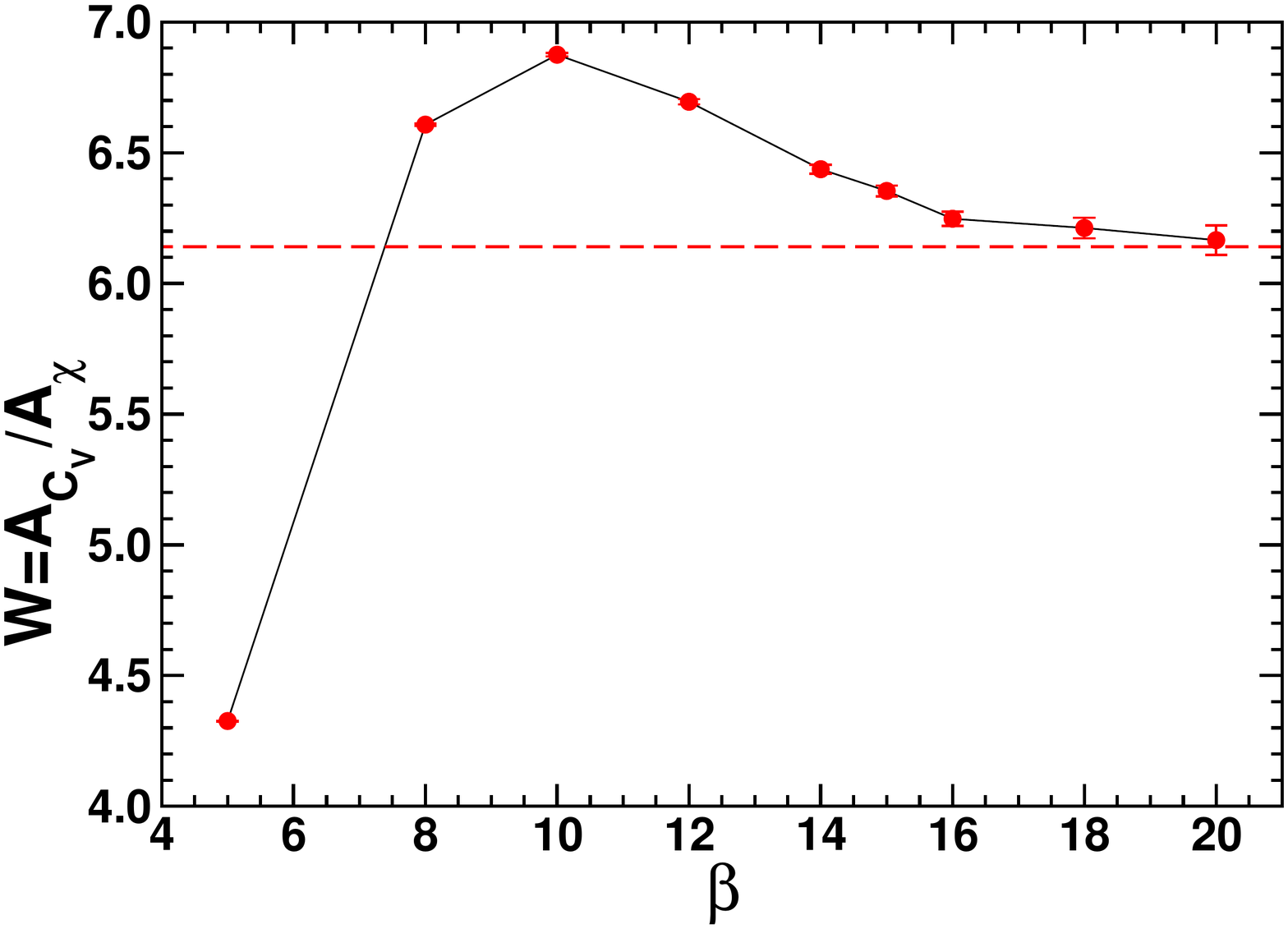}} 
\subfigure[]{\label{fig:BKappaV} \includegraphics[scale=.30]{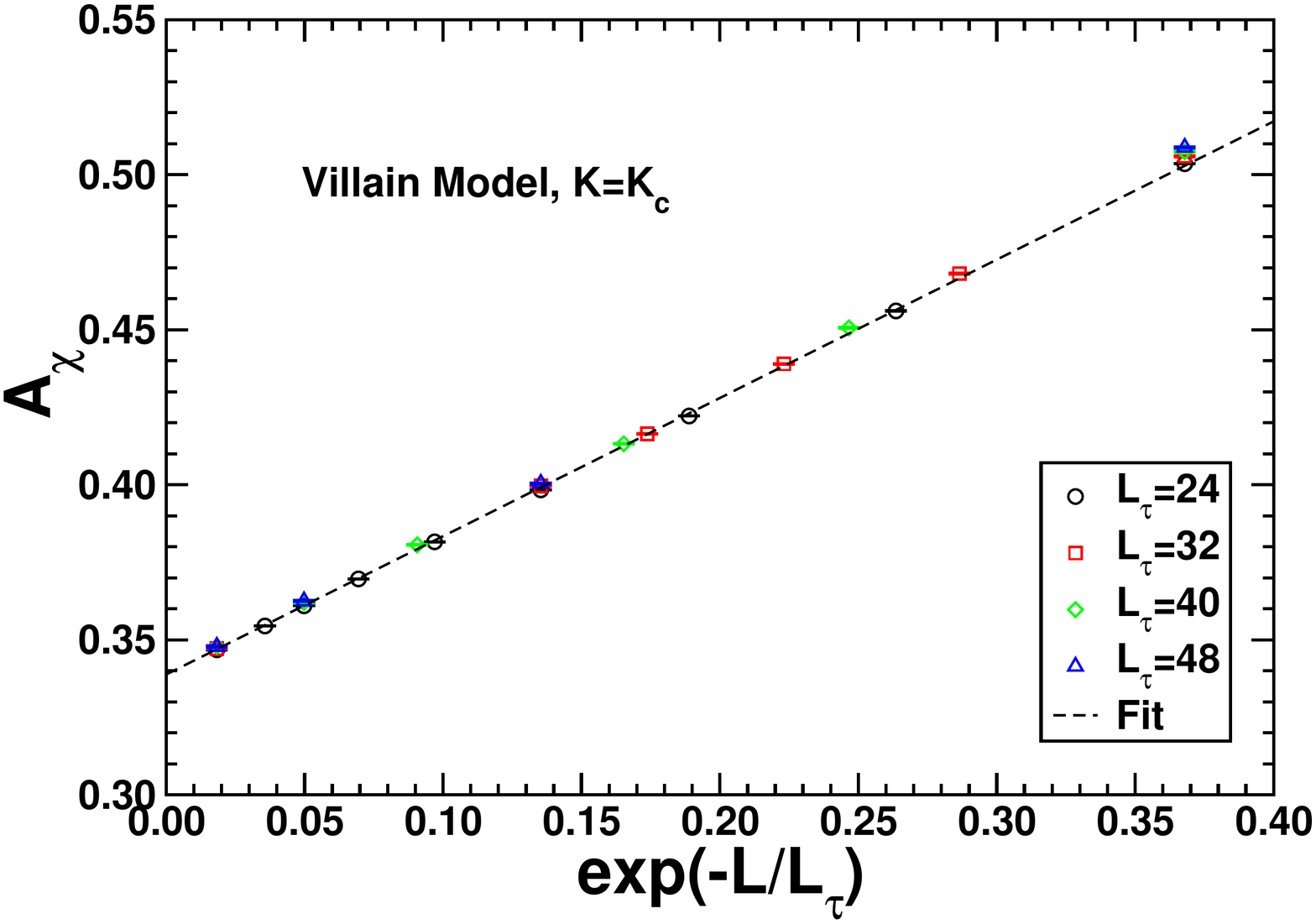}} 
\caption{\label{fig:qmc2} {\bf Quantum Monte Carlo data} 
    {\bf (a)} Scaling plot of successive crossings between $L$ and $L'$ of the scaled stiffness $L\rho$ for the quantum rotor model. Results are shown for crossing between two lattice sizes $L=30,40$ 
  and larger lattice sizes $L'$ all with $\mu=0$.
    {\bf (b)} $\beta\chi U^2$ as a function of $\beta$ at the QCP ($\mu=0$) for the quantum rotor model. All results are in the limit $L\gg\beta$.
      {\bf (c)} The ratio $W=A_{c_V}/A_\chi=\beta c_V/\chi$ at the QCP ($\mu=0$) for the quantum rotor model. The red dashed line shows the analytical estimate $A_{c_V}/A_\chi=6.14$
All results are in the limit $L\gg\beta$.
  {\bf (d)} Finite size extrapolation of the  amplitude $A_\chi=c^2\beta\chi$ at the QCP ($\mu=0$) for the Villain model. Results are shown for a range of $L$ and $L_\tau$ approaching a well defined limit $0.339(5)$
for $L\gg L_\tau$.
} 
\end{figure} 
When simulating quantum systems one is confronted with the fact that the correlation length in the temporal direction, $\xi_\tau$
could diverge differently than the spatial correlation length, $\xi$. This defines the dynamical critical exponent, $z$, through $\xi_\tau\sim\xi^z$.
For the (2+1)D $XY$ universality class we have $z=1$. However, even though the two length scales $\xi$ and $\xi_\tau$ diverge in the same manner
as the QCP is approached they could still be quite different. This means that standard finite-size scaling forms which constrain finite size corrections
to be a function of a single argument $L/\xi$ now, close to the QCP, has to have {\it two} arguments $L/\xi$ and $\beta/\xi_\tau$ since $\beta$ is the effective
extent of the temporal direction. Often it is more convenient instead of these two arguments to use the equivalent arguments $\delta L^{1/\nu}$ and $\beta L^{-z}$.
Finite size scaling then predicts~\cite{sorensen92,Wallin1994} that for $\rho$ and $\chi$: 
\begin{equation}
\rho = L^{2-d-z}\tilde\rho(\delta L^{1/\nu},\beta L^{-z}), \ \ 
\chi = L^{2-d-z}\tilde\chi(\delta L^{1/\nu},\beta L^{-z}), \ \ 
\label{eq:fss}
\end{equation}
Here, $\delta$ is the distance
to the critical point and $\tilde\rho$, $\tilde\chi$  are universal functions. In our case $z=1$,  $d=2$ and we see that 
$L\rho$ and $L\chi$ are {\it universal} and {\it independent} of $L$ at the QCP if $\beta L^{-z}$ is kept constant. This fact is exploited
in Fig.~2b of the main text to determine the location of the QCP for the quantum rotor model where curves for different $L$ are 
plotted at fixed $\beta L^{-z}$ 
showing a well defined crossing at the QCP when
$(t/U)$ is varied. 
If one analyzes the data in Fig.~2b of the main text very carefully one realizes that the curves for different $L$ do not cross {\it exactly} in a single
point. Instead, the crossing between data for size $L$ and $L'$ seem to shift to progressively higher values of $t/U$ as $L'$ is increased. This is a well-known
effect due to corrections to the scaling forms Eq.~\ref{eq:fss} and as $L'$ becomes large enough the crossings eventually converge to a single point. Assuming that 
these corrections to scaling have a power-law form it is possible to correct for them~\cite{Landau}. Our results are shown in \rfig{Critpoint} for $L=30,40$ and 
$L'=30,40,50,60,70,80,90$ with $\beta=L$ and $\Delta\tau=0.1$. As can be seen both curves point to approximately the same location of the critical point. Our final estimate for the QCP of the quantum rotor model
using is then:
\begin{equation}
(t/U)_c=0.17437(1).
  \end{equation}
In Figs.~\ref{fig:BKappaQR} and  \ref{fig:WQR} we show results for $\beta\chi U^2$ and $W=A_{c_V}/A_\chi$  for
the quantum rotor model while \rfig{BKappaV} shows the finite size extrapolation of $A_\chi$ for the Villain model.
These panels illustrate the underlying data for the estimates of these quantities in the main text. 

To facilitate the comparison to the analytical results we need to extrapolate to the thermodynamic limit, $L\to\infty$. This was done either by
directly extrapolating results for several different lattice sizes assuming finite size corrections of the form
$e^{a L}/L^\alpha,$
as well as by simulating in the zero winding sector~\cite{batrouni,pollet} for a single system with $L>L_\tau$. 
The latter procedure works well due to the fact that the main effect of increasing the lattice size is to suppress winding
number fluctuations in the spatial direction. The dominant exponential dependence of the corrections naturally arises from
the finite constant $L_\tau,\beta$ and typically one finds 
$a\sim 1/L_\tau$ (or $a \sim 1/\beta$ QR model).

The final $T\to 0$ extrapolation for the conductivity is performed in a manner analoguous to Ref.~\onlinecite{pollet} assuming corrections to the $T=0$ form
of the conductivity are powers of $\omega_n$.  For the Villain model we use the following form:
\begin{equation}
\sigma(n,L_\tau) = \sigma^{T=0}(n)-a\omega_n^w+b\omega_n^{2w}
\end{equation}
with $n$ the Matsubara index and $a,b$ constants determined in the fit. The corrections arise from the leading irrelevant operator at the
QCP with scaling dimension $w$~\cite{GZ,jzj}. Leaving $w$ a free parameter in our fits we find $w=0.877(2)$. 
As a consistency check we have verified that it is possible to obtain
largely identical results for the final $T\to 0$ extrapolated conductivity by assuming an exponential form of the corrections 
\begin{equation}
\sigma(n,L_\tau) = \sigma^{T=0}(n)-a \omega_n^we^{-b\omega_n}.
\end{equation}
In this case we obtain good fits with $w=0.887(3)$ and $a,b$ fitted constants. 
For the QR model several irrelevant operators are present, however, we are still able to extrapolate our data
assuming a very similar power-law form for the corrections.

\section{Quantifying the small frequency conductivity} 
As mentioned in the main text, in section III.B, the zero-frequency conductivity $\s(0)$
is strictly-speaking ill-defined (infinite) due to long-time tails. A useful measure of the
small-frequency $\s$ was discussed in Ref.~\onlinecite{trivedi}:
\begin{align}
  \s_*:= \int_0^{2\pi T}\frac{d\w}{2\pi T} \re \s(\w/T)\,,
\end{align}
which is nothing more than the average of the real part of the conductivity over $0\leq \w\leq 2\pi T$.
Note that this remains finite even in the presence of a logarithmic divergence of $\s$ from long-time tails.
Using the best holographic fit to the QMC data, we find $\s^*/\s_Q=0.45$, which compares well to the value
found in recent QMC simulations by the authors of Ref.~\onlinecite{trivedi}, $0.4$. Refering to Fig.~4c of the main text, we see that $\s_*$
integrates roughly over the first peak of $\s(\w/T)$, as the first inflexion point occurs near $\w=2\pi T$.

We also compute the value of $\s_*$ for the conformal fixed point of the O$(N)$ model in the $N\ra\infty$ limit 
using the exact expression\cite{damle} for $\s(\omega/T)$ and found $\s_*/\s_Q=0.59$, which 
is roughly consistent with the above.
The $1/N$ correction to the entire function $\s(\omega/T)$ is not known, so we cannot
``rigorously'' improve the field theory estimate. However, we can incorporate the known $1/N$
corrections in a makeshift fashion.
First, we recall that the $N=\infty$ conductivity can be written as\cite{damle} 
$\re \s(\w/T) =\re\s_{\rm I}(\w/T)+\re\s_{\rm II}(\w/T)$, where $\re\s_{\rm I}$ is a delta-function, 
and $\re\s_{\rm II}(\w/T)$ is non-zero only for frequencies greater than $\sim 2T$. 
The leading order $1/N$ corrections to $\s$ at small\cite{book,will-mit} and large\cite{cha} frequencies are known, 
and allow us to write down an approximate form:
\begin{align}\label{eq:approx_sig}
 \re\s(\w/T)\Big|_{N<\infty}\approx \frac{\s_0}{1+(\w\tau)^2}+(1-8\eta_\phi/3)\re\s_{\rm II}(\w/T)\,,
\end{align}
where $\eta_\phi=8/(3\pi^2 N)$ is the leading order correction to the anomalous dimension
of the rotor field. The first term corresponds to the broadening\cite{book,will-mit} of the delta-function
into a Lorentzian, with $\s_0\propto N$ and $\tau\propto N/T$ (for the prefactors, see Ref.~\onlinecite{ws}). 
The factor in front of $\s_{\rm II}$ ensures that the conductivity matches the known $1/N$ result\cite{cha}
for $\re\s(\infty)$. Performing the integral for $\s_*$ using \req{approx_sig} yields
$\s_*/\s_Q=0.43$, which lies much closer to the Monte Carlo results. 

\section{Conformal field theories}
\label{sec:cft}

The CFT describing the QCP of the BHM is the Wilson-Fisher fixed point of the field theory with imaginary time action \cite{book}
\beq
\mathcal{S}_\psi = \int d^2 x d \tau \left[ |\partial_\tau \psi|^2 + |\nabla_x \psi|^2 + s |\psi|^2 + u |\psi|^4 \right],
\eeq
where $\psi$ is the superfluid order parameter, and the QCP is at some $s=s_c$ with a superfluid phase for $s<s_c$
and an insulator for $s>s_c$. This CFT is not a gauge theory, and so it is not immediately clear \cite{sungsik12} that
it can be mapped onto a holographic dual on AdS$_4$. However, it is known\cite{dasgupta} 
that there is a particle-vortex dual of this
CFT, and an alternative description is provided by the Abelian-Higgs CFT described by 
\beq
\mathcal{S}_\phi = \int d^2 x d \tau \left[ |(\partial_\tau - i A_\tau ) \phi|^2 + |(\nabla_x - i \vec{A}) \phi|^2 + \widetilde{s}\, |\phi|^2 
+ \widetilde{u}\, |\phi|^4 \right],
\eeq
where $\phi$ is the vortex creation operator, and $A_\mu = (A_\tau, \vec{A})$ is an emergent U(1) gauge field whose flux represent the particle current of the original BHM. Now the QCP is at some $\widetilde{s}=\widetilde{s}_c$ with the insulating phase present
for $\widetilde{s} < \widetilde{s}_c$, while the superfluid appears for $\widetilde{s} > \widetilde{s}_c$.
In this form, the QCP is indeed described by a deconfined conformal gauge theory, and so a suitable matrix 
large-$N$ limit can be expected to holographically map onto a smooth AdS$_4$ geometry.\cite{sungsik12}

The above interpretation is validated by the value of $\gamma$ obtained by our analysis of the QMC data, which
positive and just below the upper bound \cite{myers11} of $\gamma = 1/12$.
In Ref.~\onlinecite{suvrat}, $\gamma$ was computed for the conformal gauge theory $\mathcal{S}_\phi$ using a 
vector large-$N$ limit in which the field $\phi$ had $N$ components: the leading order value for the particle current,
$ \epsilon_{\mu\nu\lambda} \partial_\nu A_\lambda/(2 \pi)$, was $\gamma = 1/12$.

\section{Universal ratios}
We examine the ratios of different observables in the scaling regime, and compare the
values obtained using quantum Monte Carlo, with field theory and holography.
\subsection{Charge sector}  
First, we introduce the dimensionless ratio of the charge conductivity at $\omega/T\ra\infty$
to the compressibility $\chi$:
\begin{align}
  \zeta_\infty = \frac{\s(\infty)}{\chi}\frac{T}{\hbar c^2} \,, 
\end{align}
which can be simplified to $\zeta_\infty=\hbar\s(\infty)/A_\chi$, where $\chi=A_\chi k_BT/(\hbar c)^2$.
We find
\begin{align}
  \zeta_\infty &= 0.169\,, \;\; \text{Villain model} \nn\\  
        &= 0.17\,, \;\; \text{O}(2) \text{ model, from } 1/N \text{ expansion} \nn\\
        &= 0.24\,, \;\; \text{supersymmetric Yang-Mills} \label{eq:zeta_inf-results}
\end{align}
We see that the result of our simulations agrees very well with the large-$N$ expansion
of Ref.~\onlinecite{CSY}. We mention that in the large-$N$ limit, the O$(N)$ value is\cite{CSY}
$\zeta_\infty = 0.18$ (both $\s(\infty)$ and $\chi$ stay finite as $N\ra\infty$.)  
 The value for the supersymmetric Yang-Mills gauge theory in two
spatial dimensions, $\zeta_\infty=3/(4\pi)=0.24$, is also reasonably close. Including the effects of $\g$
(which only contributes to $\chi$) leads to $\zeta_\infty=0.23$ when using $\g=\gam$, the value obtained 
from the fit to the quantum Monte Carlo data. This is slightly closer to the numerical data than the value for
super-Yang-Mills.  

Interestingly, it was suggested\cite{kovtun-ritz} that the related quantity
\begin{align}\label{eq:kr-bound}
  \zeta_0=\frac{\s(0)}{\chi}\frac{T}{\hbar c^2}\geq \frac{d+1}{4\pi(d-1)}
\end{align}
is bounded from below for CFTs, where $d$ is the spatial dimension, $d>1$. Note that this ratio involves
the d.c.\ conductivity $\s(0)$ rather than $\s(\infty)$. Focusing on two spatial
dimensions, we first observe that such a ratio will only
be meaningful in the absence of long-time tails, which make $\s(0)$ infinite as mentioned in the main text,
as well as in Appendix~\ref{sec:ltt}.  
This is the case for the supersymmetric Yang-Mills theory in the large-$N$ limit, whose conductivity 
$\s(\omega/T)$ is in fact frequency independent, such that $\zeta_\infty=\zeta_0=3/(4\pi)$,  
saturating the bound. We note that the deformation of such a theory by the four-derivative
term parametrized by $\g$ does not respect the bound when $\g<0$. This can be deduced from two facts:
$\chi$ is a monotonously decreasing function\cite{myers11} of $\g$ and $\g=0$ already saturates the bound.
Indeed, a Taylor expansion yields: $\zeta_0=(3/4\pi)(1+6\g)+\mc O(\g^2)$. 

One can ask whether $\zeta_\infty$ satisfies the bound \req{kr-bound}. However, our results \req{zeta_inf-results}
suggest that this is not the case, since both the Monte Carlo and field theory values violate \req{kr-bound}. 
It can be easily shown that $\zeta_\infty$ for the holographic theories with $\g<0$ also violates the bound.

\subsection{Mixed ratio}
In the main body, we have defined the dimensionless ratio
\begin{align}
  W=\frac{c_V}{k_B T\chi}\, ,
\end{align}
which compares the total number degrees of freedom as measured by $c_V$ to the
charge degrees of freedom ($\chi$). Partially repeating the results found in the main body:
\begin{align}
  W &= 6.2\,, \;\; \text{quantum rotor model}\,; \nn\\ 
        &= 6.14\,, \;\; \text{O}(2) \text{ model, from } 1/N \text{ expansion}\,; \nn\\
        &= 110\,, \;\; \text{supersymmetric Yang-Mills}\,. 
\end{align}
The last value is in fact $W=(3/2)(4\pi/3)^3\approx 110$. As $N\ra\infty$, $W\ra\infty$ for 
the O$(N)$ model since $c_V\sim N$
whereas $\chi\sim N^0$; this is in contrast to the superconformal-Yang-Mills CFT mentioned above,
although both are ``large-$N$'' theories.
We mention \emph{en passant} that $W$ could potentially be bounded from below
since one expects the total number degrees of freedom to exceed the charge carrying ones. This is in line with a
suggestion\cite{kovtun-ritz} that was put forward for the ratio $\eta e^2/(\s T^2)$,
where $\eta$ is the shear viscosity.  

$W$ for the superfluid-insulator transition differs by more than an order of magnitude from that
of the supersymmetric Yang-Mills theory. Such a discrepancy is not very surprising as different
conformal fixed points need not have a comparable number of charge degrees of freedom relative to the
total number. In contrast, the ratio $\zeta_\infty$ studied in the previous subsection measures the ratio
of central charges within the charge sector, and such a quantity is expected to vary less between different
(correlated) conformal fixed points. 

\section{Long-time tails} \label{sec:ltt}

The so-called long-time tails correspond to the power-law decay at large times of certain correlation functions.
This classical phenomenon has been long-known: Ref.~\onlinecite{pomeau-rev} provides an early overview for example.
In two spatial dimensions, long-time tails lead to the divergence of transport coefficients, and as such, of 
hydrodynamics itself. In this section, we examine their magnitude at the superfluid-insulator QCP, and in 
the holographic framework. 

\subsection{At the superfluid-insulator QCP} 
We estimate the strength of the long-time tail contribution to the conductivity, which we call $\ltt$, for  
the CFT describing the superfluid-insulator QCP.
A general expression that can be used to estimate $\ltt$ is given in Ref.~\onlinecite{kovtun-rev}; 
it is obtained by considering thermal fluctuations 
about 2+1D relativistic hydrodynamics. It reads 
\begin{align}
  \ltt(\w)/\s_Q=\frac{2\pi T \chi}{8\pi w (D+\g_\eta)}\ln{\frac{(D+\g_\eta)\La^2}{|\w|}} \,,
\end{align} 
where $\chi,D$ are the compressibility (charge susceptibility) and diffusion constants, respectively, while $w=\e+p$ is the
enthalpy density, which is proportional to the entropy density, $s$: $w=sT$. 
$\g_\eta=\eta/w$ is the shear viscosity normalized by $w$, so that $\g_\eta=\eta/(sT)$. Finally, $\La$
represents the cutoff beyond which the hydrodynamic description ceases to apply, i.e.\ $\La\sim T$.  
The extra factor of $2\pi$ comes from $\s_Q$ and setting $\hbar=1$. 
The expression simplifies to 
\begin{align}
  \ltt/\s_Q = \frac{T\chi}{2 c_V(TD+\eta/s)}\ln{\frac{(TD+\eta/s)\La^2}{T|\w|}}  \,.
\end{align}  
We have determined the value of the ratio $W=c_V/(k_BT\chi)$ for the quantum rotor model as discussed in the main text
and have found $W=6.2$. We can estimate the diffusion constant using the Einstein relation  
$D=\s(0)/\chi=0.47/(2\pi\times 0.339T)=0.22/T$ where  
we have used the values obtained from the simulations on the Villain model. We do not have an estimate for the
ratio of the shear viscosity to the entropy density, as this quantity is difficult to compute using either QMC
or field theoretic methods. We shall assume that it is slightly greater than the Kovtun-Son-Starinets bound\cite{kss} of $1/4\pi$,
say $1/2\pi$, although multiplying this number by 2 or $1/2$ does not lead to a big difference in the  
final answer. Combining all the numbers we obtain our final estimate:
\begin{align}
  \ltt/\s_Q= 0.21 \ln{\frac{0.38 \La^2}{T|\w|}} \,.  
\end{align}
This should be compared with $\s(\w\sim 0)/\s_Q=0.47$ obtained from the fit to our simulations. 
For $\ltt/\s_Q$ to reach $0.47$, 
it would require that $|\w|/T\sim 10^{-2}$, assuming $\La= T$. 
We thus see that the long-time tail $\ltt$ is potentially weak at the QCP under study, further justifying its neglect.   

\subsection{In holography} 
Long-time tails are suppressed in the large $N_c$ limit of the supersymmetric Yang-Mills CFTs with classical
gravity duals. Let us take an example in two spatial dimensions with $\mc N=8$ supersymmetry
introduced in Ref.~\onlinecite{abjm}, and discussed above. 
For that gauge theory, which has gauge group SU$(N_c)$ where $N_c$ is assumed to 
tend to infinity, we have $\chi,c_V$ both of order $N_c^{3/2}$, while 
$D,\eta/s$ scale like $N_c^0$. This leads to 
\begin{align}
  \s\sim N_c^{3/2}\,, \qquad \ltt \sim N_c^0\,,
\end{align}
so that the long-time tails correction subleads the leading order term by $N_c^{3/2}$. It was shown\cite{saremi} that incorporating  
$1/N_c$ corrections, which correspond to quantum corrections in the dual AdS picture, one recovers 
the long-time tails described above.  

In a holographic description which supplements that of the supersymmetric theories with higher order derivative terms, 
such as the $\g$-term used in the main body, the long-time tails also do not make their appearance. 
This is expected since such  
higher derivative corrections are classical rather than quantum.   
 
 
%

\end{document}